\documentclass[aps,prc,twocolumn,nofootinbib,amsmath,superscriptaddress,floatfix,longbibliography]{revtex4-2}%showpacs
\usepackage{graphicx,epsfig,longtable}
\usepackage{mathptmx}
\usepackage[light,first]{draftcopy} % ,bottomafter
\usepackage{epstopdf}
\usepackage[pdftex]{hyperref}
\usepackage{wasysym}
\usepackage{multirow}
\usepackage{eqnarray,amsmath,amsbsy}
\usepackage{booktabs}
\usepackage{appendix}
\usepackage{xcolor}

\begin{document}
%0
\newcommand{\ks}[1]{\textcolor{green}{#1}}
\newcommand{\ig}[1]{\textcolor{red}{#1}}
\newcommand{\sg}[1]{\textcolor{blue}{#1}}

\preprint{APS/123-QED}

\title{Photoneutron reactions on $^{165}$Ho and $^{169}$Tm in the giant dipole resonance region}
% Force line breaks with \\

\author{I.~Gheorghe}\email{ioana.gheorghe@nipne.ro}
\affiliation{National Institute for Physics and Nuclear Engineering,
Horia Hulubei (IFIN-HH), 30 Reactorului, 077125 Bucharest-Magurele, Romania}

\author{S.~Goriely} %stephane.goriely@ulb.be
\affiliation{Institut d'Astronomie et d'Astrophysique, Universit\'e Libre de Bruxelles, Campus de la Plaine, CP-226, 1050 Brussels, Belgium}

\author{K.~Stopani} %kstopani@sinp.msu.ru
\affiliation{Lomonosov Moscow State University, Skobeltsyn Institute of Nuclear Physics, 119991 Moscow, Russia}

\author{S.~Belyshev} %belyshev@depni.sinp.msu.ru
\affiliation{Lomonosov Moscow State University, Faculty of Physics, 119991 Moscow, Russia}

\author{M. Krzysiek} % mateusz.krzysiek@ifj.edu.pl 
\affiliation{National Institute for Physics and Nuclear Engineering,
Horia Hulubei (IFIN-HH), 30 Reactorului, 077125 Bucharest-Magurele, Romania}
\affiliation{Institute of Nuclear Physics Polish Academy of Sciences, PL-31342 Krakow, Poland}

\author{D.~Filipescu} %dan.filipescu@nipne.ro
\affiliation{National Institute for Physics and Nuclear Engineering,
Horia Hulubei (IFIN-HH), 30 Reactorului, 077125 Bucharest-Magurele, Romania}

\author{H.~Wang} %wanghw@sari.ac.cn
\affiliation{Shanghai Advanced Research Institute, Chinese Academy of Sciences, No.99 Haike Road, Zhangjiang Hi-Tech Park, 201210 Pudong Shanghai, China}

\author{G.~Fan} %fangongtao@zjlab.org.cn
\affiliation{Shanghai Advanced Research Institute, Chinese Academy of Sciences, No.99 Haike Road, Zhangjiang Hi-Tech Park, 201210 Pudong Shanghai, China} 

\author{L.~Liu} %liulongxiang@sinap.ac.cn
\affiliation{Shanghai Advanced Research Institute, Chinese Academy of Sciences, No.99 Haike Road, Zhangjiang Hi-Tech Park, 201210 Pudong Shanghai, China} 

\author{H.~Scheit} %hscheit@ikp.tu-darmstadt.de
\affiliation{Institut f\"ur  Kernphysik, Technische Universit\"at Darmstadt, Darmstadt, 64289, Germany}

\author{D.~Symochko} %dsymochko@ikp.tu-darmstadt.de
\affiliation{Institut f\"ur  Kernphysik, Technische Universit\"at Darmstadt, Darmstadt, 64289, Germany}
\affiliation{Framatome GmbH, 91056 Erlangen, Germany}

\author{T.~Aumann} %taumann@ikp.tu-darmstadt.de
\affiliation{Institut f\"ur  Kernphysik, Technische Universit\"at Darmstadt, Darmstadt, 64289, Germany}
\affiliation{GSI Helmholtzzentrum f\"ur Schwerionenforschung, 64291 Darmstadt,Germany}
\affiliation{Helmholtz Forschungsakademie Hessen f\"ur FAIR (HFHF), GSI Helmholtzzentrum f\"ur Schwerionenforschung, 64291 Darmstadt, Germany}

\author{T.~Renstr\o{}m} %therese@xal.no
\affiliation{Department of Physics, University of Oslo, N-0316 Oslo, Norway}
\affiliation{Expert Analytics AS, N-0179 Oslo, Norway}

\author{G.M.~Tveten} %gry@xal.no
\affiliation{Department of Physics, University of Oslo, N-0316 Oslo, Norway}
\affiliation{Expert Analytics AS, N-0179 Oslo, Norway}

\author{F.~Furmyr} % fridafu@gmail.com  
\affiliation{Department of Physics, University of Oslo, N-0316 Oslo, Norway}

\author{A.~G\"orgen} % andreas.gorgen@fys.uio.no
\affiliation{Department of Physics, University of Oslo, N-0316 Oslo, Norway}
\affiliation{Norwegian Nuclear Research Centre, N-0316 Oslo, Norway}

\author{A.C.~Larsen} %a.c.larsen@fys.uio.no
\affiliation{Department of Physics, University of Oslo, N-0316 Oslo, Norway}
\affiliation{Norwegian Nuclear Research Centre, N-0316 Oslo, Norway}

% \author{J.E.~Midtb\o{}} % JorgenEriksson.Midtbo@fhi.no  jorgenem@gmail.com
% \affiliation{Department of Physics, University of Oslo, N-0316 Oslo, Norway}

\author{S.~Siem} % 	sunniva.siem@fys.uio.no
\affiliation{Department of Physics, University of Oslo, N-0316 Oslo, Norway}
\affiliation{Norwegian Nuclear Research Centre, N-0316 Oslo, Norway}

\author{V.~Orlin} %
\affiliation{Lomonosov Moscow State University, Skobeltsyn Institute of Nuclear Physics, 119991 Moscow, Russia}

\author{T.~Ari-izumi} %ariizumi.takashi@qst.go.jp
\affiliation{Konan University, Department of Physics, 8-9-1 Okamoto, Higashinada, Kobe 658-8501, Japan}

\author{S.~Miyamoto} %miyamoto.shuuji.ile@osaka-u.ac.jp %shujimiyamoto@icloud.com
\affiliation{Laboratory of Advanced Science and Technology for Industry, University of Hyogo, 3-1-2 Kouto, Kamigori, Ako-gun, Hyogo 678-1205, Japan}
\affiliation{Institute of Laser Engineering, Osaka University, 2-6 Yamadaoka, Suita, Osaka 565, Japan}

\author{H. Utsunomiya} %\email{hiro@konan-u.ac.jp}
\affiliation{Konan University, Department of Physics, 8-9-1 Okamoto, Higashinada, Kobe 658-8501, Japan}

%\affiliation{Shanghai Advanced Research Institute, Chinese Academy of Sciences, No.99 Haike Road, Zhangjiang Hi-Tech Park,  201210 Pudong Shanghai, China}

\date{\today}% It is always \today, today,

\begin{abstract}

Photoneutron reactions were investigated for the deformed $^{165}$Ho and $^{169}$Tm nuclei from the vicinity of the neutron emission threshold up to $\sim$40~MeV, well above the giant dipole resonance (GDR) region, using quasimonochromatic laser Compton scattering $\gamma$-ray beams provided at the NewSUBARU facility, Japan. A high-and-flat efficiency moderated array of $^3$He counters was used for the neutron detection and an associated neutron multiplicity sorting method for extracting the $(\gamma,\,1nX)$, $(\gamma,\,2nX)$, $(\gamma,\,3nX)$ and $(\gamma,\,4nX)$ reaction cross sections and average neutron emission energies. The present $^{165}$Ho cross sections were compared to existing data, revealing discrepancies with the Saclay multiplicity sorting results and an overall 10$\%$ strength difference with the Livermore ones. There are no other data for $^{169}$Tm. The total photoneutron cross sections $\sigma(\gamma,\,Sn)$ were determined by summing the partial $(\gamma,\,inX)$ components. GDR parameters based on phenomenological Lorentzian models were extracted by fitting the present $\sigma(\gamma,\,Sn)$ data with adjustments for the missing contribution of charged-particle-only reactions not observed experimentally. For both nuclei we observed high energy structures at 20-25~MeV, matching giant quadrupole resonance predictions from the microscopic Kombined Model of Fotonuclear Reactions (KMFR). Based on the present centroid energies of the first and second GDR peaks, hydrodynamic model predictions gave intrinsic electric quadrupole moments of +7.00(34)~b and +7.38(28)~b for the ground states of $^{165}$Ho and $^{169}$Tm, respectively. The present experimental excitation functions and photoneutron energies were compared to statistical model calculations. Using the EMPIRE statistical model code, we performed a sensitivity test to phenomenological models of photon strength functions and nuclear level densities. The TALYS code was used to reproduce the present experimental $(\gamma,\,inX)$ cross sections and average neutron energies using microscopic nuclear level density models. 

\end{abstract}

%\pacs{Valid PACS appear here}
\maketitle

% \tableofcontents

\section{Introduction} \label{sec_intro}

The present measurement of $(\gamma,\,in)$ reaction cross sections and average neutron emission energies in $^{165}$Ho and $^{169}$Tm is part of a comprehensive photoneutron study in the giant dipole resonance (GDR) region~\cite{gheorghe_2017,filipescu_2024,gheorghe_2024,gheorghe_2025,gheorghe_2026} with a flat efficiency neutron detector and quasi-monochromatic $\gamma$-ray beams at the NewSUBARU facility~\cite{amano_2009,horikawa_2010}. The project is motivated by the need for accurate and reliable photonuclear data for fundamental nuclear research purposes, as well as for a variety of current or emerging applications, such as medical, energy and astrophysics~\cite{kawano_2020}. 

\begin{figure*}[t]
\centering
\includegraphics[width=0.8\textwidth, angle=0]{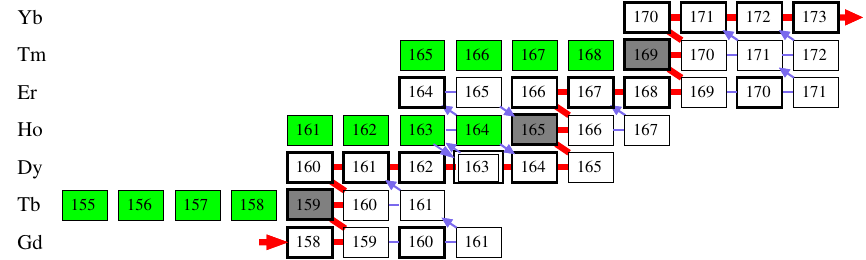}
\caption{The $s$-process nucleosynthesis path in the Z=64 to 70 atomic number region (thick solid boxes indicate stable, and thin solid boxes unstable, isotopes). The thick red lines show the main $s$-process path and the thin blue ones the minor branchings. Grey boxes show the target nuclei for the present ($^{165}$Ho and $^{169}$Tm) and our recent study ($^{159}$Tb~\cite{gheorghe_2026}), while green boxes show the nuclei populated by the investigated $(\gamma,\,in)$ reactions. }\label{fig_s_process_crop}     
\end{figure*}

In particular for heavy-mass nuclei, where the photoabsorption cross section is well approximated by the total photoneutron one, $(\gamma,\,in)$ data are used to directly extract the photon strength function (PSF) at excitation energies above the neutron emission threshold ($S_n$)~\cite{goriely_2019}, as well as to determine the E1 dipole polarizability and the symmetry energy~\cite{goriely_2020}. Photoabsorption cross sections are used to estimate quadrupole electric moments for the ground states of deformed target nuclei and also provide information on their deformation~\cite{danos_1958}, which is particularly interesting for rare earth nuclei recently revealed to be characterized by triaxiality~\cite{Otsuka_2025}.

Besides determining the PSF in the $^AX$ target nucleus, the simultaneous reproduction of the $(\gamma,\,in)$ photoneutron cross sections and average neutron emission energies through statistical model calculations generates a coherent set of experimentally constrained PSFs and nuclear level densities (NLDs) in the $^{A-i}X$ isotopic chain, where $i$ cycles from 1 to 4 in the present study. This leads to improved predictions of neutron captures and photo-disintegrations in the investigated isotopic chains, which are of general interest for nucleosynthesis calculations of the rapid, intermediate and slow neutron capture processes ($r$-, $i$- and $s$-processes) and of the $p$-process essentially made of $(n,\,\gamma)$ and $(\gamma,\,n)$, $(\gamma,\,p)$ and $(\gamma,\,\alpha)$ photodisintegrations of a large network of stable and unstable nuclei~\cite{arnould_2003,arnould_2020}.

Figure~\ref{fig_s_process_crop} shows the reaction path of the $s$-process in the area between Gd and Yb, highlighting the nuclei populated in the present $^{165}$Ho and $^{169}$Tm study and in the recent $^{159}$Tb~\cite{gheorghe_2026} one performed at the same laboratory. We notice that improved predictions for $(n,\,\gamma)$ cross sections in the Tb, Ho and Tm isotopes are particularly important for $s$-process calculations~\cite{Kaeppeler_2011} in the represented area, especially for the unstable, and thus experimentally challenging, $^{163,166,167}$Ho and $^{170,171}$Tm branching-point nuclei. Cross sections for neutron captures on the $^{163}$Ho branching point, which is interestingly populated by the bound-state $\beta$-decay of $^{163}$Dy that becomes beta-active under stellar conditions, may influence an $s$-process contribution to the production of the $p$-nucleus $^{164}$Er and also determine the remaining $^{163}$Ho abundance at the end of the $s$-process~\cite{Jaag_1996,Hayakawa_2008}. Also required is the improved modeling of the $(n,\,\gamma)$ cross sections on the unstable $^{166,167}$Ho, as it was recently shown that they can influence the $^{166,167}$Er abundances in high neutron densities conditions~\cite{Pogliano_2023}, in particular during the $i$-process \cite{Martinet_2024}. The $^{170}$Tm and $^{171}$Tm are major and minor branching points, respectively, for which the $(n,\,\gamma)$ cross sections affect the abundances of $^{170-172}$Yb~\cite{Reifarth_2003,Guerrero_2020}.

Cross sections of photoneutron reactions on $^{165}$Ho in the GDR region were measured in the 1960's using quasi-monochromatic $\gamma$-ray beams from positron annihilation in flight at the Saclay and Livermore facilities. However, the $^{165}$Ho data sets disagree, with the Saclay $(\gamma,\,1n)$ cross sections being significantly larger than the Livermore results. Since the $(\gamma,\,2n)$ cross sections are in agreement, a uniform scaling factor cannot be applied to reconcile the two data sets. This issue is general for the Saclay and Livermore data, for which characteristic systematic discrepancies between the two facilities remain unresolved~\cite{wolynec_1987,varlamov_2014}. Although the $^{169}$Tm$(\gamma,\,n)$ reaction is suitable for activation measurements and could thus be used as a monitor reaction, there are no photonuclear data for $^{169}$Tm except for an unpublished $(\gamma,\,n)$ measurement at 13~MeV~\cite{Wolsey_2019}.

In the present study we investigated photoneutron reactions in $^{165}$Ho and $^{169}$Tm in the GDR region using laser Compton scattering (LCS) $\gamma$-ray beams at NewSUBARU. $(\gamma,\,in)$ reactions with $i$=1 to 4 were observed in an energy range between the neutron separation energy in each nucleus and $\sim$40 MeV. A preliminary version of the present photoneutron cross sections was provided to the International Atomic Energy Agency (IAEA) Coordinated Research Project (CRP) on Updating the Photonuclear Data Library~\cite{kawano_2020}. Here we report updated cross sections using the latest methods of characterizing the LCS $\gamma$-ray beam spectral density~\cite{filipescu_2023} and flux~\cite{utsunomiya_2018}, and of neutron emission multiplicity sorting~\cite{gheorghe_2021}. We also complete the experimental data set with new quantities: the average energies of $(\gamma,\,in)$ neutron emission spectra.

\begin{figure*}[t]
\centering
\includegraphics[width=0.9\textwidth, angle=0]{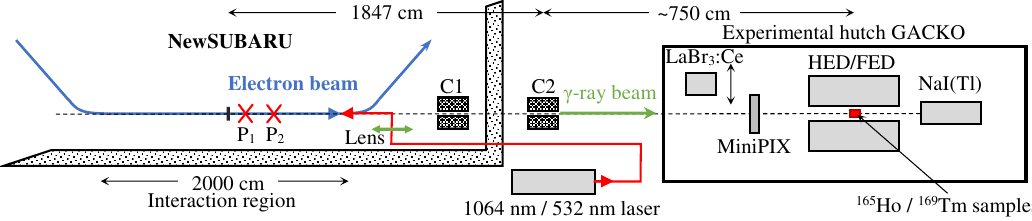}
\caption{Schematic layout (not to scale) of the BL01 LCS $\gamma$-ray beamline and GACKO experimental hutch at NewSUBARU. C$_1$ and C$_2$ mark the 10~cm long lead collimators (3 and 2~mm aperture, respectively), while P$_1$ and P$_2$ are the focal point positions for the 1064~nm and the 532~nm lasers, respectively~\cite{filipescu_2023,hashimoto_2022}.}\label{fig02_BL01_PRC_Ho165_Tm169_crop}
\end{figure*}

In Sec.~\ref{sec_exp_method}, we present the experimental technique with focus on characteristics of the incident LCS $\gamma$-ray beams. The data analysis methods are discussed in Sec.~\ref{sec_data_analysis}. Results are discussed and compared with preceding data in Sec.~\ref{sec_results} and with theoretical calculations in Sec.~\ref{sec_STAT_calc}. A summary and conclusions are given in Sec.~\ref{sec_summary}.

\section{Experiment} \label{sec_exp_method}

In this study, we performed 1- to 4-fold coincidence measurements of the neutrons emitted in photonuclear reactions on $^{165}$Ho and $^{169}$Tm to determine absolute photoneutron cross sections and average energies of photoneutron emission spectra in the GDR excitation energy range. Since the experimental method has been published in detail~\cite{gheorghe_2017,filipescu_2024,gheorghe_2024,gheorghe_2025}, only a general description is given here, complemented with the specific features of the present measurement. 

The experiment was conducted at the LCS $\gamma$-ray beamline BL01 of the NewSUBARU electron storage and accelerator ring~\cite{amano_2009,horikawa_2010} using the GACKO experimental hutch, which are both represented schematically in Fig.~\ref{fig02_BL01_PRC_Ho165_Tm169_crop}. The $\gamma$-ray beams of 8 to 43~MeV maximum energy were produced by scattering 1064 and 532~nm wavelength laser photons on relativistic electron beams circulating in the NewSUBARU ring. The absolute energy of the $\gamma$-ray beams was set at 58 values for $^{165}$Ho and at 65 values for $^{169}$Tm by controlling the electron beam energy, which is calibrated with 10$^{-5}$ precision~\cite{utsunomiya_2014}. The LCS $\gamma$-ray beam was collimated and transported to the experimental hutch, where it impinged on metallic $^{165}$Ho and $^{169}$Tm samples (99.9~$\%$ chemical purity) placed in the center of the neutron detection array. 
The $^{165}$Ho sample took the form of a disk of either 3.518 or 3.641~$\mathrm{g/cm}^2$ ($\approx$4~mm) thickness for measurements below 26~MeV incident energy, and a stack of the two disks for the higher energy measurements.
For $^{169}$Tm a 3.775~$\mathrm{g/cm}^2$ ($\approx$4~mm) thick disk was used for the measurements below 15~MeV and a 7.401~g/cm$^2$ thick stack of two such disks was used for energies above that.
The sample diameter was of 20~mm for both isotopes, considerably larger than the $\approx$4~mm diameter beam spot of the pencil-like $\gamma$-ray beam~\cite{Ariizumi_2023}. 

The LCS $\gamma$-ray beam was delivered in narrow pulses of 40-60 ns duration. For neutron counting measurements conducted below the two-neutron separation energy $S_{2n}$~(see Table~\ref{table_Sxn}), where only the $(\gamma,\,n)$ reaction channel is open, the beam repetition rate was set at 20~kHz. Above this energy threshold, the repetition rate was reduced to 1~kHz to facilitate neutron multiplicity sorting. The incident $\gamma$-ray flux, measured using a large-volume NaI detector positioned in the beam dump and analyzed via the associated Poisson fitting method~\cite{utsunomiya_2018}, was of $\sim10^5$ $\gamma$/s for neutron counting and of $\sim10^4$ $\gamma$/s for neutron multiplicity sorting irradiations.

\begin{figure*}[t]
\centering
\includegraphics[width=0.8\textwidth, angle=0]{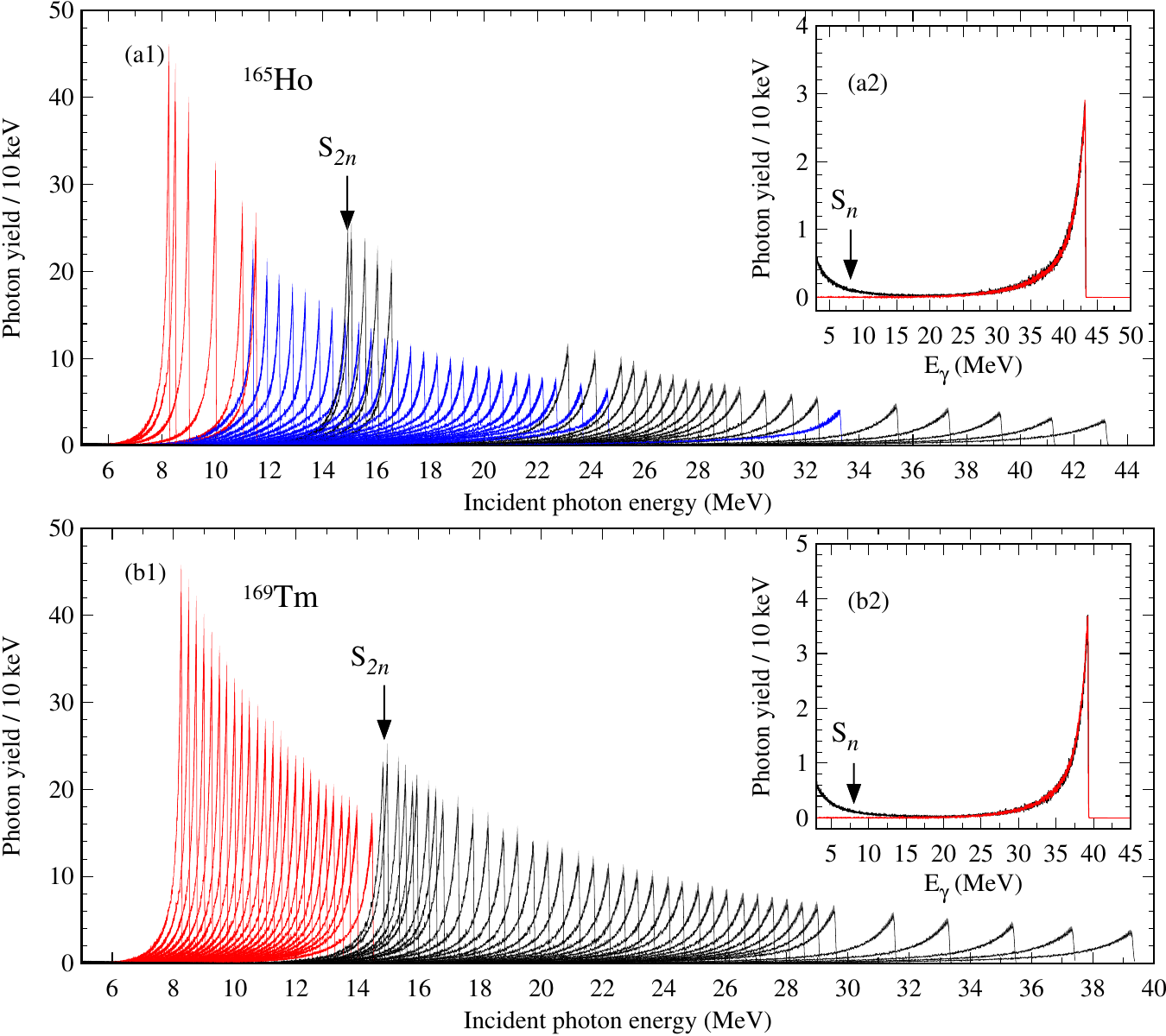} %{../ETC/MIX_PLOT_INC_SPECTRA/fig_inc_sp_mix.pdf}
\caption{Spectral distributions of LCS $\gamma$-ray beams used for the (a) $^{165}$Ho and (b) $^{169}$Tm irradiations. The spectra are normalized to have identical areas. (a1) and (b1) Spectra of $\gamma$-ray beams produced using the 1064~nm wavelength laser (red) and the 532~nm one (black and blue). Blue histograms show a set of spectra with lower energy resolution. Insets show the transmission comparison between spectral distributions of LCS $\gamma$-ray beams before (red) and after (black) passing through 8~mm thick (a2) $^{165}$Ho and (b2) $^{169}$Tm samples, where the pairs of the spectra are normalized to the maximum energy amplitude.}\label{fig_inc_sp_mix}     
\end{figure*}

Figure~\ref{fig_inc_sp_mix} shows the spectral distributions of the $\gamma$-ray beams incident on the $^{165}$Ho and $^{169}$Tm samples. The red histograms correspond to $\gamma$-ray beams produced using the low-energy 1064~nm wavelength laser and the black and blue ones to the beams produced with the 532~nm laser. The energy spectra were obtained using the $\texttt{eliLaBr}$ code~\cite{filipescu_2022,filipescu_2023} by reproducing the experimental LaBr$_3$ monitor detector response for each irradiation energy. 
For $^{169}$Tm we notice in Fig.~\ref{fig_inc_sp_mix}(b1) the well-known~\cite{gheorghe_2024,gheorghe_2025} smooth energy spread increase with the increase in the $\gamma$-ray beam maximum energy, along with a rather abrupt reduction of the energy spread when switching from the 1064~nm laser to the 532~nm laser, a feature determined by Compton scattering kinematics~\cite{filipescu_2022}. 
Thus, for the $^{169}$Tm irradiations, the $\gamma$-ray beam energy resolution increased smoothly from 1.4 to 2.1~$\%$ in full width at half maximum (FWHM) for the beams produced with the 1064~nm laser and from 1.3 to 2.8~$\%$ for the ones produced with the 532~nm laser.
However, for $^{165}$Ho, several irradiations were performed in non-optimal conditions of the laser beam optics~\cite{hashimoto_2022,filipescu_2023}, which led to the generation of $\gamma$-ray beams with energy widths slightly larger than expected. The energy spectrum of these $\gamma$-ray beams is represented in blue in Fig.~\ref{fig_inc_sp_mix}(a1). Even so, the energy spread of the $\gamma$-ray beams used for the $^{165}$Ho irradiations remained within 1.3 and 3.3~$\%$ in FWHM. 

We accounted in our data analysis for the spectral changes caused by the electromagnetic (EM) interaction between the incident $\gamma$-ray beams and the sample material. The magnitude of the spectral modification is proportional to the amount of target material and the energy of the incident $\gamma$-ray beam. Figures~\ref{fig_inc_sp_mix} (a2) and (b2) show the transmission comparison between the original (red) and the EM-altered (black) spectral distributions of the highest energy LCS $\gamma$-ray beams used in this experiment, which were 43.22 MeV and 39.32 MeV for $^{165}$Ho and $^{169}$Tm, respectively. A low-energy tail, consisting of secondary photons produced by EM interactions in the $\approx$8~mm targets, is evident in the figures. When the energy spectrum of these secondary photons exceeds the $S_n$, they contribute to the photoneutron reaction emission in the target.

The neutrons emitted in the irradiated samples were recorded by $^3$He counters (10 atm. pressure, 45~cm active length) arranged in three concentric rings around the beamline axis and embedded in a high density polyethylene moderator matrix. A $\sim$60--80$\%$ high-efficiency detection (HED) geometry described in Refs.~\cite{utsunomiya_2015,gheorghe_2025} was used for the low-energy irradiations performed using the 1064~nm laser. 
A flat efficiency detection (FED) configuration was used for the higher energy measurements. The FED array has a detection efficiency of 37.3$\%$ with a variation of 5$\%$ within the 10~keV~--~5~MeV interval of average energies of evaporation neutron spectra and is described in Ref.~\cite{utsunomiya_2017}.

For neutron counting measurements below the two-neutron emission threshold, we employed a beam-on/beam-off macro time structure (80 ms beam-on, 20 ms beam-off) to discriminate reaction neutrons from background ones. Above the two-neutron separation energy, we recorded neutron multiplicity events with 1 up to 4 neutron coincidences. A $k$-fold event corresponds to $k$ neutrons detected within a 1~ms interval between $\gamma$-ray beam bunches. Background separation was achieved by analyzing neutron arrival time histograms for each ring and coincidence order, as described in Refs.~\cite{gheorghe_2021,gheorghe_2024}. 

\begin{figure*}[t]
\centering
\includegraphics[width=0.95\textwidth, angle=0]{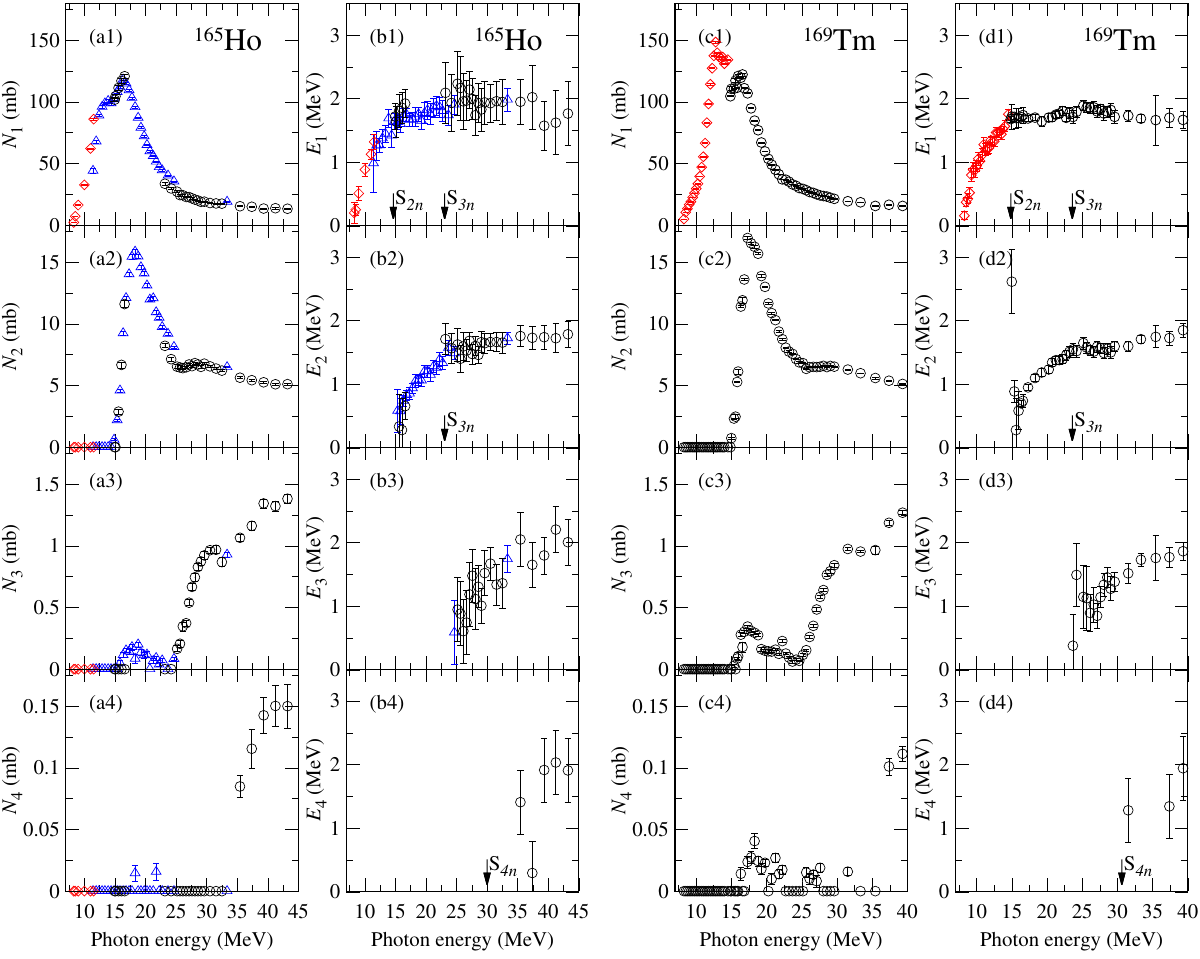} %{../ETC/MIX_PLOT_NEVRATE_EK/fig_nev_rate_mix.pdf}
\caption{Raw experimental data for (left~--~a, b) $^{165}$Ho and (right~--~c, d) $^{169}$Tm before neutron detection efficiency correction and neutron multiplicity sorting. $i$-fold neutron cross sections $N_i$ defined in Eq.~\eqref{eq_ni} are shown in columns (a) and (c) and average energies of neutrons recorded in $i$-fold coincidences $E_i$ are shown in columns (b) and (d). Red diamonds show results obtained with the 1064~nm wavelength laser, while black circles and blue triangles ones show those obtained with the 532 nm laser, using similar colors as for the photon data in Fig.~\ref{fig_inc_sp_mix}. $E_i$ error bars represent the statistical component and a 3$\%$ uncertainty in the neutron detection efficiency calibration, and the $N_i$ uncertainties are statistical only. }\label{fig_nev_rate_mix}
\end{figure*}

\section{Data analysis} \label{sec_data_analysis}

We began the data analysis by examining the detected $i$-fold neutron events without attempting to link them to the original emitted neutron multiplicity and without applying efficiency corrections. The raw experimental data sets obtained following the procedure detailed in Ref.~\cite{gheorghe_2021} are shown in Fig.~\ref{fig_nev_rate_mix}:
\paragraph{} The $i$-fold neutron cross sections $N_i$ are defined as:
\begin{equation} \label{eq_ni}
N_i (E_\mathrm m) = \cfrac{n_i(E_\mathrm m) }{N_\gamma (E_\mathrm m) n_\mathrm T \xi(E_\mathrm m) }, 
\end{equation} 
where $E_\mathrm m$ is the maximum energy of the incident LCS $\gamma$-ray beam, $n_i$ is the number of $i$-fold neutron coincidence events, $N_\gamma$ is the number of $\gamma$ photons incident on the target and $n_\mathrm T$ is the concentration of target nuclei. $\xi=[1-\mathrm{exp}(-\mu L)]/\mu$ is a thick target correction factor given by the target thickness $L$ and attenuation coefficient $\mu$.
\paragraph{} The average energy of neutrons recorded in $i$-fold coincidence events $E_i$ is obtained through the ring ratio method as described in Refs.~\cite{gheorghe_2021,filipescu_2024,gheorghe_2024,gheorghe_2025}.

The $N_i$ cross sections for $^{165}$Ho are shown in Fig.~\ref{fig_nev_rate_mix}(a1)~--~(a4) and for $^{169}$Tm in Fig.~\ref{fig_nev_rate_mix}(c1)~--~(c4). The $E_i$ values for $^{165}$Ho are shown in Figs.~\ref{fig_nev_rate_mix}(b1)~--~(b4) and for $^{169}$Tm in Fig.~\ref{fig_nev_rate_mix}(d1)~--~(d4). In red are the results obtained using $\gamma$-ray beams produced with the low energy 1064~nm wavelength laser and the HED array. The higher energy points obtained with the 532~nm laser and the FED array are shown in black and blue, where blue triangles show results obtained with the lower energy resolution LCS $\gamma$-ray beam mentioned in Sec.~\ref{sec_exp_method}. 

The error bars of the $i$-fold cross sections and of the corresponding average neutron energies are determined by the statistical fluctuations of $i$-fold neutron counts, where a 3$\%$ uncertainty in the neutron detection efficiency is also included in the $E_i$ error bars. We again notice~\cite{gheorghe_2021} that the average neutron energy results are more sensitive to statistical fluctuations than the $i$-fold cross sections. A statistical scatter is most visible for the $^{165}$Ho $E_i$ values at high excitation energies, above 25~MeV average, represented by black circles in Figs.~\ref{fig_nev_rate_mix}(b1)~--~(b4). 

For both nuclei, we note in Figs.~\ref{fig_nev_rate_mix}~(a1),~(b1),~(c1) and (d1) a discontinuity in the $N_1$ cross sections and the $E_1$ average energies at $\sim$12~--~15~MeV excitation energy (where $E_x \approx E_\gamma$ for typical LCS experiments on heavy targets such as $^{165}$Ho), when switching from using the HED detector together with the 1064 nm laser (red diamonds) to using the FED detector together with the 532 nm laser (blue and black symbols). Being generated by the different efficiencies of the neutron counter arrays and by the different energy spectra of the $\gamma$ beams obtained with the two lasers, this discontinuity is natural and present in all such measurements~\cite{gheorghe_2024,gheorghe_2025}. At higher incident energies, the discontinuities between the $^{165}$Ho $N_i$ and $E_i$ values represented by blue and black symbols are produced by probing the excitation functions and average neutron energies with LCS $\gamma$-ray beams of different spectral characteristics, as discussed in the previous section and represented in Fig.~\ref{fig_inc_sp_mix}~(a1).

\begin{table}[b]
\caption{\label{table_Sxn} $^{165}$Ho and $^{169}$Tm $i$ neutrons separation energies $S_{in}$. All values are given in MeV.}
\begin{ruledtabular}
\begin{tabular}{lcccccc}
   Nucleus &   \textrm{S$_n$} & \textrm{S$_{2n}$}& \textrm{S$_{3n}$} & \textrm{S$_{4n}$} & \textrm{S$_{5n}$} \\ \colrule
 \textrm{$^{165}$Ho}          & 7.989  &  14.66          & 23.07             & 29.99             & 38.88 \\
 \textrm{$^{169}$Tm}          & 8.034  &  14.88          & 23.60             & 30.63             & 39.74  \\ 
\end{tabular}
\end{ruledtabular}
\end{table}

The second step in the data analysis was to link the detected $i$-fold neutron events to the emitted ones, and thus obtain the $(\gamma,\,in)$ photoneutron cross sections and average energies. This was done using the statistical treatment of neutron coincidence events described in Ref.~\cite{gheorghe_2021}, which models the multiple-firing of all available combinations of $(\gamma,\,in)$ photoneutron reactions with corresponding contributions given by the reaction cross sections, concentration of target nuclei and incident $\gamma$-ray flux. The $S_{in}$ neutron separation energies with $i$ from 1 to 5 are given for $^{165}$Ho and $^{169}$Tm in Table~\ref{table_Sxn}.

As a third and last step in the data analysis, corrections were brought to the so-obtained $(\gamma,\,in)$ photoneutron cross sections and average energies by taking into account the non-monochromaticity of the incident $\gamma$-ray beams. Following the method described in Refs.~\cite{Renstrom18,LarsenTveten23}, a test function was iteratively adjusted until the folding between it and the EM-altered LCS $\gamma$-ray spectra reproduced the uncorrected cross sections and average neutron energies obtained at the previous step. Details are given in Ref.~\cite{filipescu_2023} for $(\gamma,\,in)$ cross section unfolding and in Ref.~\cite{filipescu_2024} for average neutron energy unfolding.   

For the error propagation, we first estimated the uncertainty for the photoneutron cross sections and average neutron energies before the non-monochromaticity correction. These represent the quadratic sum of the statistical uncertainties in the $i$-fold neutron detection and the uncertainties of 3$\%$ for the incident flux, 3$\%$ for the neutron detection efficiency and 0.25~-~0.5$\%$ for the target thickness. Finally, for each $\sigma_{\gamma,in}$ and $E_{\gamma,in}$ curve, the total uncertainties were estimated by adding (subtracting) the error bars to (from) the uncorrected central values and applying the non-monochromaticity correction separately to the two so-obtained data sets, as described in Refs.~\cite{Renstrom18,LarsenTveten23}.

\begin{figure*}[t]
\centering
\includegraphics[width=0.92\textwidth, angle=0]{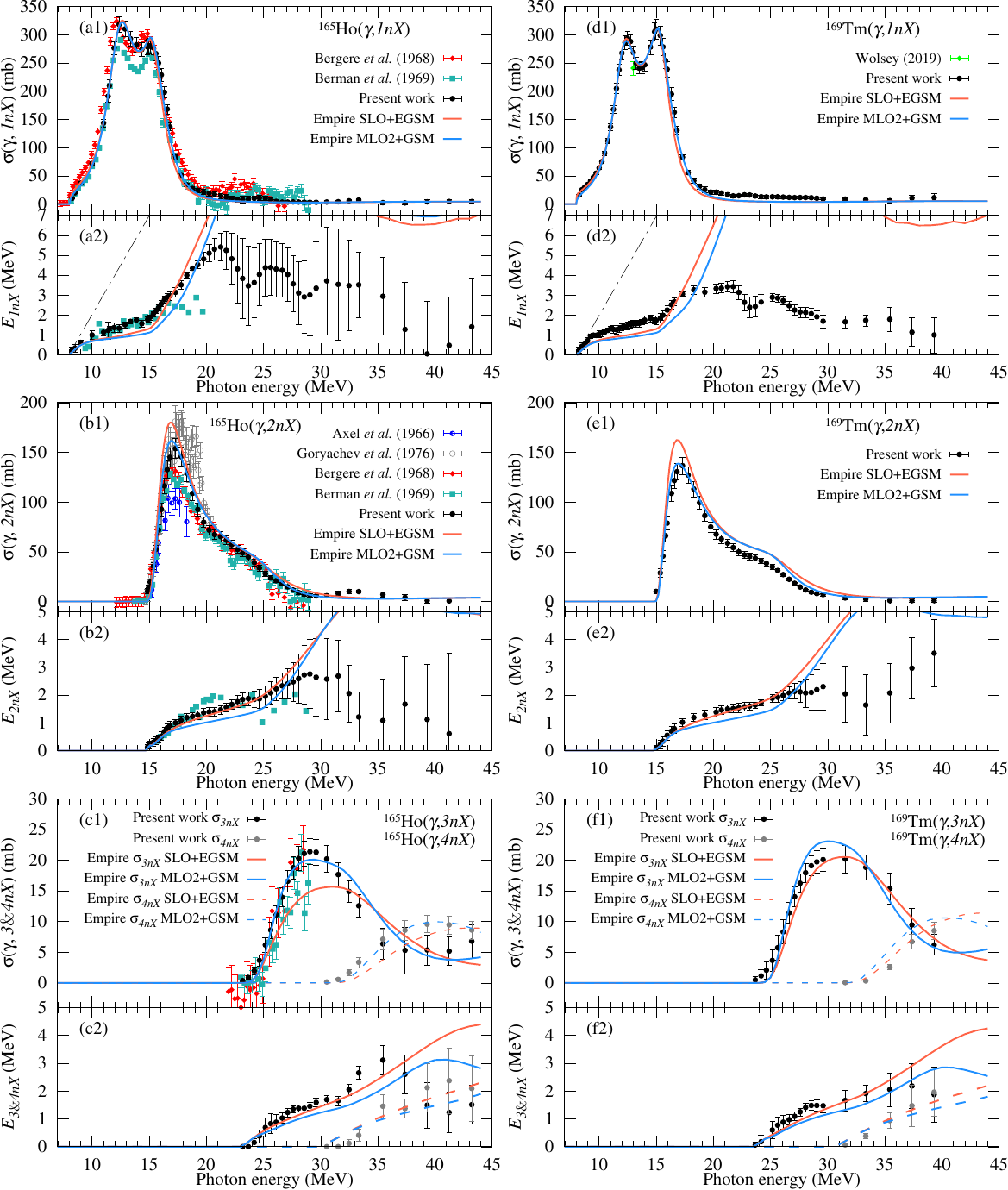} %{../ETC/MIX_PLOT_CS_UNFOLD/fig_cs_en_gxn.pdf}
\caption{Present $(\gamma,\,inX)$ experimental data for $^{165}$Ho and $^{169}$Tm compared with existing data and with EMPIRE statistical model calculations. Error bars show total uncertainties. The dash-dotted lines in (a2) and (d2) correspond to the maximum $(\gamma,\,n)$ neutron energies given by kinematics (A-1)/A$\cdot$($E_\gamma-S_n$).}\label{fig_cs_en_gxn}     
\end{figure*}

\section{Experimental results}\label{sec_results}

The present photoneutron experimental results are shown in Fig.~\ref{fig_cs_en_gxn}. As only neutron detectors were used for recording radiation emitted from the irradiated samples, no experimental discrimination could be made between reaction channels with only neutron emission and those with both neutron and charged particle emission, while reactions with only charged particle emission were not observed at all. Thus, we use the IAEA recommended notations and define the present photoneutron cross section results as the $\sigma_{inX}$ sum cross sections of all photoneutron reactions with $i$ neutrons in the final state, accompanied or not by charged particle emission:
\begin{align}
\sigma_{inX}          & \equiv \sigma_{\gamma,\,inX} \nonumber \\
                      & \equiv \sigma(\gamma,\,inX) \nonumber \\
                      & = \sigma(\gamma,\,in) + \sigma(\gamma,\,inp) + \sigma(\gamma,\,in\alpha) + \dots       
\end{align}
A similar $E_{inX}$ notation is used for the corresponding average neutron energies. 

\subsection{Photoneutron results for $^{165}$Ho}

The present $\sigma_{1nX}$, $\sigma_{2nX}$, and $\sigma_{3nX}$ photoneutron cross sections for $^{165}$Ho plotted in Figs.~\ref{fig_cs_en_gxn}~(a1), (b1) and (c1), respectively, are compared to existing data measured in the 1960s using positron annihilation in flight $\gamma$-ray beams both at the Saclay facility by Berg\`ere~\emph{et~al.}~\cite{Bergere_1968} (red diamonds) and at Livermore by Berman~\emph{et~al.}~\cite{Berman_1969} (green squares). For the $(\gamma,\,2n)$ reaction, there are also bremsstrahlung cross sections measured by Goryachev~\emph{et~al.}~\cite{Goryachev_1976} and older Saclay data of Axel~\emph{et~al.}~\cite{Axel_1966}, which are plotted in Fig.~\ref{fig_cs_en_gxn}~(b1). Figure~\ref{fig_cs_en_gxn}(c1) shows also the present $\sigma_{\gamma,\,4nX}$ cross sections, for which there are no other existing data. 

\begin{figure}[t]
\centering
\includegraphics[width=0.95\columnwidth, angle=0]{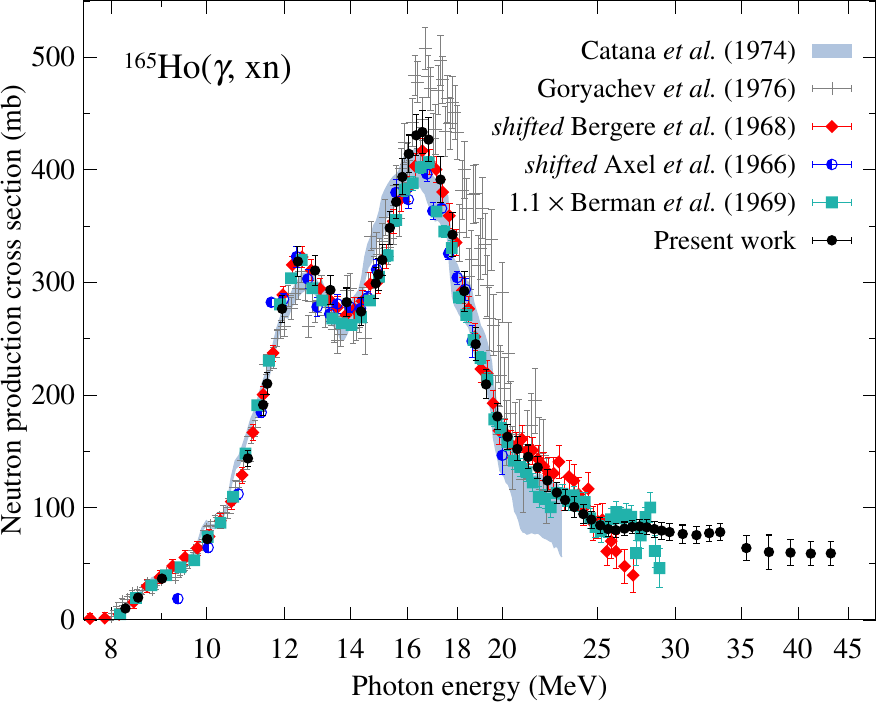} %{../ETC/Ho165_PLOT_CS_UNFOLD/fig_cs_ny.pdf}
\caption{Present total photoneutron production (yield) cross sections for $^{165}$Ho obtained as $\sigma_{xn} = \sum_i i \cdot \sigma_{inX}$ along with existing data. The Saclay data~\cite{Bergere_1968,Axel_1966} are shifted by +0.35~MeV, while the Livermore data~\cite{Berman_1969} are multiplied by a factor of 1.1, see text for details.}\label{fig_cs_ny_Ho}     
\end{figure}

\begin{figure}[t]
\centering
\includegraphics[width=0.95\columnwidth, angle=0]{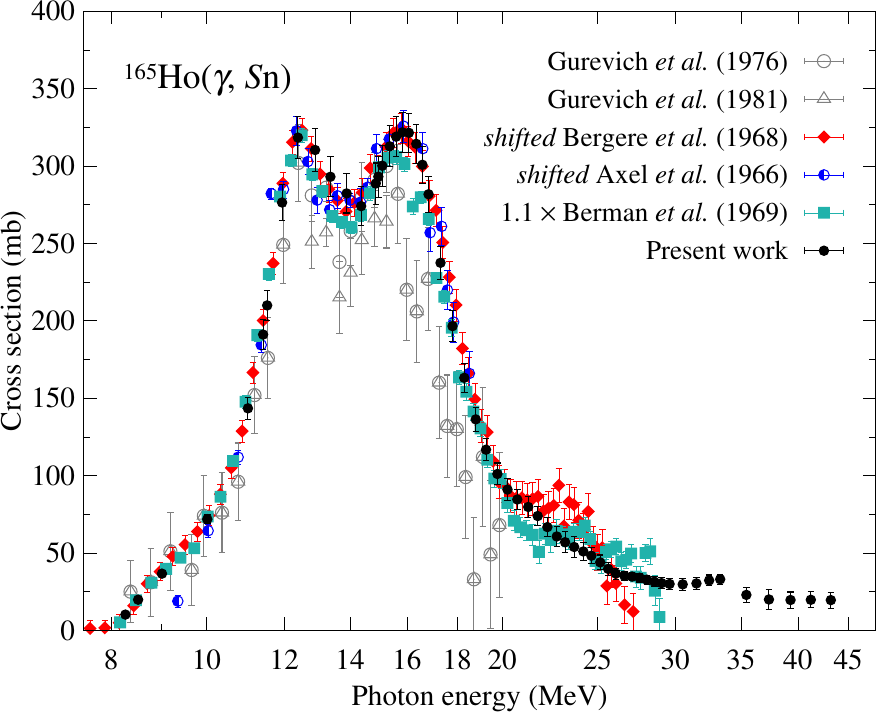} %{../ETC/Ho165_PLOT_CS_UNFOLD/fig_cs_abs.pdf}
\caption{Present total photoneutron cross sections for $^{165}$Ho obtained as $\sigma_{Sn} = \sum_i \sigma_{inX}$ along with existing data. The Saclay data~\cite{Bergere_1968,Axel_1966} are shifted by +0.35~MeV, while the Livermore data~\cite{Berman_1969} are multiplied by a factor of 1.1, see text for details.}\label{fig_cs_abs_Ho}     
\end{figure}

We first compare our data with those from Saclay, noting the following key points:
\begin{itemize}
\item An energy shift exists, evident in the low-energy $(\gamma,\,1nX)$ cross section.
\item Above $S_{2n}$, cross section strengths differ: present $\sigma_{1nX}$ are smaller than Saclay ones, $\sigma_{2nX}$ are larger.
\end{itemize}
In order to understand the origin of these discrepancies, we examine Saclay's primary data: the total photoneutron production (yield) cross section $\sigma_{xn}$ is defined as:
\begin{equation}
\sigma_{xn}  = \sum_i i \cdot \sigma_{\gamma,\,inX}.
\end{equation}
In the Saclay and Livermore experiments, the $\sigma_{xn}$ cross section is the primary physical quantity from which the partial $\sigma_{inX}$ cross sections are extracted by neutron multiplicity sorting. In the present experiment, $\sigma_{inX}$ are the primary quantities.

Figure~\ref{fig_cs_ny_Ho} compares the present $\sigma_{xn}$ results with those from Saclay and with the other data available in the literature. 
In order to obtain a good agreement between the position of the GDR centroids observed in the Saclay experiment with those observed at NewSUBARU and Livermore, we find that the Saclay data have to be energetically shifted by +0.35~MeV, as represented in Fig.~\ref{fig_cs_ny_Ho}. We note that +0.27~MeV shift has been previously proposed by Wolynec and Martins~\cite{wolynec_1987} from the analysis of the Saclay and Livermore data. Relatively good agreement is also obtained with the GDR peak positions observed in the bremsstrahlung experiments of Catan\u{a}~\emph{et~al.}~\cite{Catana_1974} and of Goryachev~\emph{et~al.}~\cite{Goryachev_1976}. Moreover, this shift yields a very good agreement between the strength of Saclay and present $\sigma_{xn}$ cross-sections over the wide energy range between $S_n$ and 20~MeV. Similar agreement holds for the total photoneutron cross section $\sigma_{Sn}$ defined as
\begin{equation}
\sigma_{Sn}  = \sum_i \sigma_{\gamma,\,inX}
\end{equation}
and shown in Fig.~\ref{fig_cs_abs_Ho}.

The good agreement between the energetically shifted $\sigma_{xn}$ and $\sigma_{Sn}$ Saclay cross sections and the present results indicates that the discrepancies between the strengths of the partial $(\gamma,\,1nX)$ and $(\gamma,\,2nX)$ cross sections in the energy range between $S_{2n}$ and 20~MeV originate from the neutron multiplicity sorting method. More precisely, it suggests a possible misassignment of $(\gamma,\,2nX)$ events as $(\gamma,\,1nX)$ ones in the Saclay analysis, which would lead to artificially increased values for the $(\gamma,\,1nX)$ cross sections and respectively decreased values for the $(\gamma,\,2nX)$ cross section. This confirms the conclusion of Wolynec and Martins~\cite{wolynec_1987}, who first addressed the issue for several nuclei ranging from $^{89}$Y to $^{238}$U and, based on independent $(e,\,n)$ and $(e,\,2n)$ cross sections for $^{181}$Ta, suggested a faulty neutron multiplicity sorting in the Saclay measurements.

One discrepancy remains between the shifted Saclay cross sections and the present one: the extra strength observed at Saclay at excitation energies between 20 and 25~MeV is considerably higher than the low energy shoulder observed in the present and Livermore $\sigma_{xn}$ and $\sigma_{Sn}$ results. However, given the specific difficulty of positron annihilation in flight experiments, namely the high neutron background in measurements at energies above the GDR, we won't overinterpret this.

We move on to comparing the present $^{165}$Ho photoneutron cross section results with the Livermore ones. Figures~\ref{fig_cs_en_gxn}(a1), (b1) and (c1) show that the Livermore partial $\sigma_{inX}$ cross sections are systematically below the NewSUBARU ones. Again, we investigate the primary Livermore data, namely the total photoneutron production (yield) cross section $\sigma_{xn}$. We find that, by multiplying the Livermore $\sigma_{xn}$ data by a factor of 1.1, a good overlap is obtained with the present and the energy shifted Saclay results, as shown in Fig.~\ref{fig_cs_ny_Ho}. A similar agreement between the NewSUBARU, the Saclay data shifted by +0.35~MeV and the Livermore data increased by 10~$\%$ is observed in Fig.~\ref{fig_cs_abs_Ho} for the total photoneutron cross section $\sigma_{Sn}$, where only the bremsstrahlung results of Gurevich~\emph{et~al.}~\cite{Gurevich_1976,Gurevich_1981} are systematically below the other measurements.

The $^{165}$Ho $E_{1nX}$, $E_{2nX}$, and the $E_{3nX}$ along with the $E_{4nX}$ average neutron energies are shown in Figs.~\ref{fig_cs_en_gxn}(a2), (b2) and (c2), respectively. The $E_{inX}$ average energies continuously increase starting from the corresponding $S_{in}$ reaction thresholds up to $\sim$5~MeV excitation energy above the following $S_{(i+1)n}$ threshold, from where they start a slow decrease. The present $E_{1nX}$ and $E_{2nX}$ energies are compared with the Livermore values in Figs.~\ref{fig_cs_en_gxn}(a2) and (b2) respectively. For both reactions the two data sets are in agreement only on the low excitation energy region between the reaction threshold and the opening of the next reaction channel. The increase of the present $E_1$ energies becomes faster starting at $S_{2n}$, while the Livermore energies maintain their slope. Although less evident, this is also observed in the $E_2$ energies. The average neutron energies are determined with best precision on the increasing slope. The large error bars associated with the average neutron energies on the high excitation energy decreasing slope are generated by the cross section decrease along with the opening of reaction channels with higher neutron emission multiplicities.

\subsection{Photoneutron results for $^{169}$Tm}

The present $\sigma_{1nX}$, $\sigma_{2nX}$, and $\sigma_{3nX}$ along with the $\sigma_{4nX}$ photoneutron cross sections for $^{169}$Tm are plotted in Figs.~\ref{fig_cs_en_gxn}(d1), (e1) and (f1), respectively. Besides the unpublished $\sigma_{1nX}$ measurement at 13~MeV excitation energy of Wolsey~\cite{Wolsey_2019} which is plotted in Fig.~\ref{fig_cs_en_gxn}(d1), there are no other existing photoneutron data on $^{169}$Tm. 

The present photoneutron cross sections for $^{169}$Tm exhibit characteristics similar to those observed for $^{165}$Ho. 
The $(\gamma,\,1nX)$ reaction cross section displays two distinct peaks and a subsequent decline at energies exceeding the GDR, with broad structures present at excitation energies of approximately 20-25~MeV in both $^{165}$Ho and $^{169}$Tm. 
The $(\gamma,\,1nX)$ and $(\gamma,\,2nX)$ cross sections reach maxima of approximately 300~mb and 150~mb, respectively, before decreasing to near-zero values at excitation energies above 30 MeV. The $(\gamma,\,3nX)$ and $(\gamma,\,4nX)$ reaction cross sections reach maxima of 20~mb and 10~mb, respectively, for both $^{169}$Tm and $^{165}$Ho.

The $E_{1nX}$, $E_{2nX}$, and $E_{3nX}$ along with the $E_{4nX}$ average photoneutron energies for $^{169}$Tm are plotted in Figs.~\ref{fig_cs_en_gxn}(d2), (e2) and (f2), respectively. 
The average energy of neutrons emitted in the $^{169}$Tm$(\gamma,\,1nX)$ reactions increases from the reaction threshold up to excitation energies of 17~MeV, followed by a gradual decline. 
The maximum $E_{1nX}$ average neutron energy for $^{169}$Tm is 3.4~MeV, significantly lower than the 5.4~MeV maximum observed for $^{165}$Ho. 
At low excitation energies, from $S_n$ and up to 9~MeV, high energy neutrons are emitted in the $(\gamma,\,1nX)$ reaction, approaching the kinematically determined maximum. 
This behavior is present in $^{169}$Tm, not in $^{165}$Ho, but has also been observed in the case of $^{209}$Bi~\cite{gheorghe_2017} and $^{208}$Pb~\cite{gheorghe_2024}.

\begin{figure*}[t]
\centering
\includegraphics[width=0.95\textwidth, angle=0]{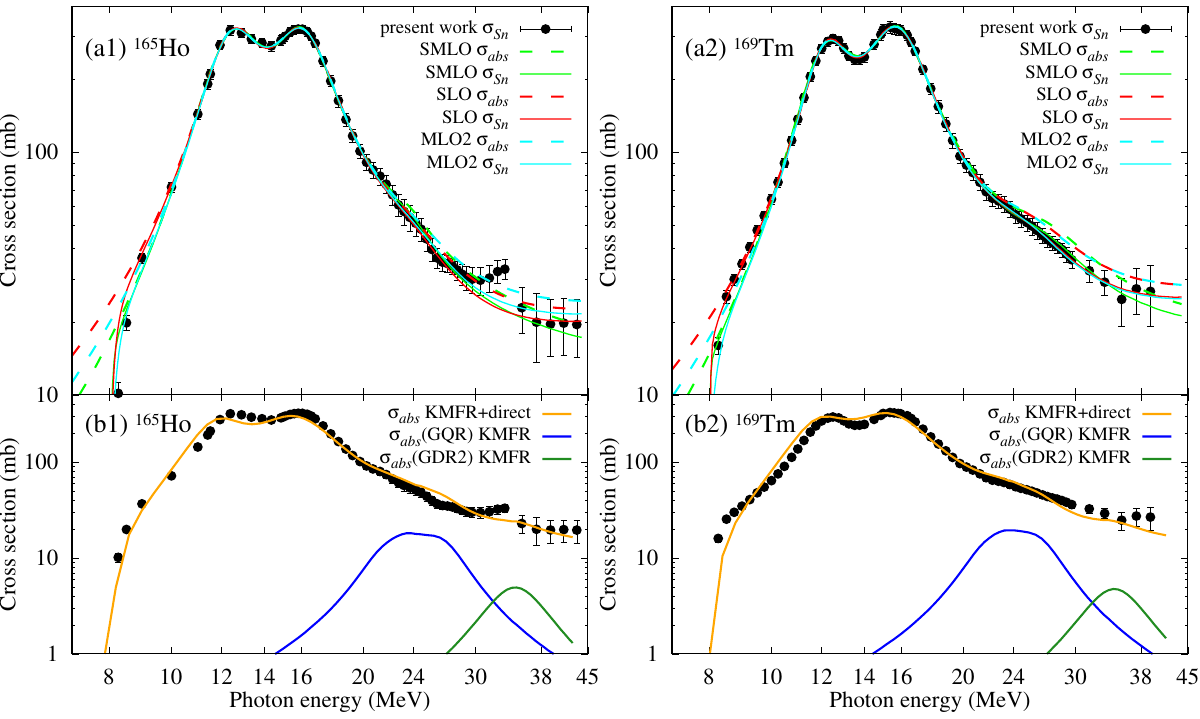} %{../ETC/MIX_PLOT_CS_UNFOLD/fig_abs.pdf}
\caption{Present total photoneutron cross sections obtained as $\sigma_{Sn} = \sum_i \sigma_{inX}$ for (left) $^{165}$Ho and (right) $^{169}$Tm compared with (upper panels) Lorenzian fits and (lower panels) microscopic predictions of the KMFR model.}\label{fig_abs}     
\end{figure*}

\section{Theoretical interpretation} \label{sec_STAT_calc}

Here we first examine the photoabsorption excitation functions of the two nuclei. The experimental data are fitted using phenomenological Lorentzian parameterizations and also compared to microscopic calculations in order to explain the nature of the structures observed above the GDR range. Statistical model calculations are performed using the EMPIRE~\cite{herman_2007_empire} and TALYS~\cite{koning_2023_talys} codes to reproduce the experimental $(\gamma,\,inX)$ cross sections and average neutron energies. Finally, based on the Lorentzian GDR fit, the electric quadrupole moment is extracted.  

\subsection{Photoabsorption cross sections} \label{sec_gabs_calc}

Based on the fact that, in the investigated energy range, neutron emission is the dominant mode of decay for the photon excited states, we extracted the photoabsorption theoretical fit parameters from the present experimental total photoneutron cross sections. For this, we described the photoabsorption cross section as a sum of the terms corresponding to the GDR excitation given by $\sigma_\mathrm{GDR}$, and the quasi-deuteron (QD) contribution $\sigma_\mathrm{QD}$~\cite{capote2009_ripl,plujko_2018}:  
\begin{equation}
\sigma_{abs}(E_\gamma) = \sigma_\mathrm{GDR}(E_\gamma) + \sigma_\mathrm{QD}(E_\gamma),
\end{equation}
where by $\sigma_\mathrm{QD}$ we refer to the cross section for the photodisintegration of a neutron-proton pair, and $E_\gamma$ is the incident photon energy. 

\subsubsection{Fit of the photoabsorption excitation function}

Figures~\ref{fig_abs}(a1) and (a2) show the present experimental $\sigma_{Sn}$ data for $^{165}$Ho and $^{169}$Tm, respectively, fitted with three Lorentzian models of the PSF: the Standard Lorentzian (SLO)\cite{capote2009_ripl}, the Simple Modified Lorentzian (SMLO)~\cite{plujko_2018} and the Modified Lorentzian (MLO2)~\cite{capote2009_ripl} plus the QD contribution. Two Lorentzians were used to describe the absorption in the 10 to 20~MeV excitation energy range, where the GDR in the deformed $^{165}$Ho and $^{169}$Tm is distinctly split into two components. A third Lorentzian was used to describe the structure observed for both nuclei as a low shoulder at excitation energies above 20~MeV. Table~\ref{table_GDR_par} gives for each nucleus the fit values of the Lorentzian centroids, widths and peak cross sections and the normalization factor for the QD photo-absorption cross section used within the EMPIRE code. 

\begin{table*}[t]
\caption{\label{table_GDR_par} Lorentzian parameters obtained by fitting the present total photoneutron cross sections with the SLO~\cite{capote2009_ripl}, SMLO~\cite{plujko_2018} and MLO2~\cite{capote2009_ripl} functions. The fit was corrected for the missing charged-particle-only component, see text for details. $QD$ stands for the normalization factor for the quasi-deuteron photoabsorption cross section used within the EMPIRE code. }
\begin{ruledtabular}
\begin{tabular}{llcccccccccr}
    \textrm{CN}          & Lorentzian & E(1)           & $\Gamma$(1)      & $\sigma$(1)      & E(2)           & $\Gamma$(2)      & $\sigma$(2)      & E(3)           & $\Gamma$(3)      & $\sigma$(3)  & $QD$   \\ 
                         & type       & \textrm{(MeV)} & \textrm{(MeV)}   & \textrm{(mb)}    & \textrm{(MeV)} & \textrm{(MeV)}   & \textrm{(mb)}    & \textrm{(MeV)} & \textrm{(MeV)}   & \textrm{(mb)}&      \\ \colrule
    \textrm{$^{165}$Ho}  & SLO        & 12.53          & 2.90             & 263.6            & 16.03          & 4.16             & 270.3            & 23.00          &  8.51            & 14.43        & 1.81 \\      %  2025-12-03      SLO  ratio_GDR=   1.440
    \textrm{$^{165}$Ho}  & MLO2       & 12.57          & 3.05             & 268.8            & 16.06          & 4.05             & 260.6            & 23.01          & 10.81            & 15.59        & 1.78 \\      %  2025-12-03      MLO2 ratio_GDR=   1.255
    \textrm{$^{165}$Ho}  & SMLO       & 12.68          & 3.40             & 283.0            & 16.19          & 3.84             & 247.7            & 23.52          &  4.70            &  9.60        & 0.66 \\      %  2025-12-03      SMLO ratio_GDR=   0.949   
    \textrm{$^{169}$Tm}  & MLO2       & 12.30          & 2.57             & 227.5            & 15.76          & 4.15             & 284.3            & 25.01          & 12.00            & 22.07        & 2.03 \\      %  2025-12-03      MLO2 ratio_GDR=   1.933
    \textrm{$^{169}$Tm}  & SLO        & 12.28          & 2.38             & 222.0            & 15.72          & 4.25             & 292.0            & 25.18          & 12.00            & 20.82        & 2.19 \\      %  2025-12-03      SLO  ratio_GDR=   2.277  
    \textrm{$^{169}$Tm}  & SMLO       & 12.37          & 2.91             & 241.1            & 15.89          & 4.02             & 274.5            & 26.70          & 10.83            & 16.00        & 1.00 \\      %  2025-12-03      SMLO ratio_GDR=   1.506 
\end{tabular}
\end{ruledtabular}
\end{table*}

As in the case of $^{159}$Tb~\cite{gheorghe_2026}, it was necessary to adjust the fit parameters of the photoabsorption cross section by taking into account also the strength of reactions with emission of charged particles unaccompanied by neutrons. 
Thus, for each Lorentzian model, Figs.~\ref{fig_abs}(a1) and (a2) show separately the $\sigma_{abs}$ fit and the corresponding $\sigma_{Sn}$ total photoneutron cross section, where the latter reproduces the experimental data. The EMPIRE code predicts a 3$\%$ energy integrated contribution in the 8 to 44~MeV energy range for the reaction channels with charged particle emission unaccompanied by neutrons in both of the investigated nuclei. Their contribution becomes visible above $\sim$20~MeV excitation energy, where the $\sigma_{abs}$ fit is significantly above the $\sigma_{Sn}$ calculations and experimental data. 

We notice that all three PSF parametrizations describe the GDR well around the two peaks, with almost similar results for the centroid energies. However, differences are observed at lower energies, around and below the neutron separation threshold. For both nuclei, the SMLO parameterization yields the smallest cross sections for the low-energy region among the three fits, while the SLO parameterization yields the largest. Notably, for $^{169}$Tm, the SLO and MLO2 fits underestimate the cross section in the energy region up to 10 MeV. 

\subsubsection{Assignment of giant quadrupole resonance structure}

Differences in photoabsorption descriptions using the three PSF models are also evident at energies above the GDR peaks, partly due to the challenge of accurately reproducing the structure at 20-25 MeV through fitting.  
In previous experimental measurements on $^{165}$Ho \cite{Bergere_1968,Berman_1969}, a considerable bump with irregular shape in the $(\gamma, Sn)$ cross section was observed, constituting approximately an additional 140 mb~MeV on the high-energy tail of the GDR curve at 20--26 MeV. In the theoretical analysis \cite{Ligensa_1966}, this bump was interpreted in terms of the hydrodynamic model as a broad fine-structured giant quadrupole resonance (GQR) with a total area of about 160 mb~MeV. It should be noted that the details in the fine structure of the bump are different in the measurements \cite{Bergere_1968,Berman_1969} and are hard to separate from noise. In the present measurement a smooth kink is visible at these energies. This kink is taken into account by fitting it with the third Lorentzian curve in the photoabsorption cross section (see Table~\ref{table_GDR_par}) with an integrated area of 193 mb~MeV in $^{165}$Ho.

To better understand the nature of the high energy structures observed in the experiment, we used the Kombined Model of Fotonuclear Reactions (KMFR) \cite{Ishkhanov_2015,kmfr_website}, in which the photoabsorption process is described in a semi-microscopic way. Figures~\ref{fig_abs}(b1) and (b2) show the present experimental $\sigma_{Sn}$ compared to the KMFR predictions, where we notice the good reproduction of the GDR photoabsorption cross section in the peak region. The calculated GQR cross sections are also shown by the blue lines. The calculated parameters of the GQR for $^{165}$Ho and $^{169}$Tm as well as for $^{159}$Tb and $^{197}$Au which were studied in previous NewSUBARU experiments~\cite{gheorghe_2025,gheorghe_2026} are compared to experimental ones in Table~\ref{table_GQR_par}.  
Deformation splitting results in the formation of three components of the GQR with different absolute values of the angular momentum projection $|M|$. For comparison with the experimental data, the ``total'' values of the GQR parameters were obtained by fitting the sum of three components with a single Lorentzian curve. It is seen that the KMFR predictions are more or less close to each other and present an averaged behaviour of the observed GQR resonances of the four nuclei. In this picture, the energy spread of the 20--26~MeV bump in $^{165}$Ho results from the deformation splitting of the GQR, and a very similar resonance is predicted in similarly deformed $^{169}$Tm. It is tempting to attribute the fact that the observed GQR width in $^{169}$Tm (12.00~MeV) is 1.4~times as large as in $^{165}$Ho to the uncertainty of the fitting procedure of a weak bump superimposed on a background. A strong GQR is expected by KMFR in $^{197}$Au, but only a small peak was observed experimentally. After averaging, the theoretically calculated centroid energy of the GQR follows very closely the $135\cdot A^{-1/3}$ approximation~\cite{bohr_mottelson}. 

An additional structure can be seen in the experimental photoabsorption cross section for $^{165}$Ho and $^{169}$Tm at about 34 MeV. An overtone of the giant dipole resonance (GDR2) formed from $3\hbar\omega$ single-particle transitions is expected at this energy. A simplified description of this overtone in KMFR without the effect of deformation predicts for both nuclei a broad GDR2 peak shown by the green lines in Fig.~\ref{fig_abs}(b1) and (b2). However, its $\Gamma \approx 8$ MeV and integrated cross section $\sigma_\text{int} \approx 75$ mb~MeV are too small to justify its identification in the experimental data.

\begin{table}[t]
\caption{\label{table_GQR_par} GQR peak parameters: experimental values of centroid energy $E_\text{exp}$ and energy-integrated cross section $\sigma_\text{int,exp}$ obtained in the present and previous~\cite{gheorghe_2025,gheorghe_2026} NewSUBARU studies and KMFR predictions of $E$, $\sigma_\text{int}$ and the GQR width $\Gamma$ for deformation-splitting components of the GQR corresponding to different absolute values of the angular momentum projection $|M|$. Energy and width values are given in MeV, and integrated cross sections in mb~MeV.}
\begin{tabular}{lcccccc}
\hline
Nucleus & $|M|$ & $E$ & $\Gamma$ & $\sigma_\text{int}$ & $E_\text{exp}$ & $\sigma_\text{int,exp}$ \\
\hline
$^{159}$Tb  &     0 &  21.80 & 4.11 & 32.89  &            &           \\
            &     1 &  23.65 & 4.60 & 77.38  &            &           \\
            &     2 &  26.69 & 5.43 & 98.61  &            &           \\
            & Total &  24.55 & 7.39 & 232.72 &   24.8     &  291.82   \\
$^{165}$Ho  &     0 &  21.55 & 4.04 & 34.13  &            &           \\
            &     1 &  23.39 & 4.52 & 80.37  &            &           \\
            &     2 &  26.42 & 5.33 & 102.56 &            &           \\
            & Total &  24.20 & 7.33 & 223.61 &   23.2     &   192.89  \\
$^{169}$Tm  &     0 &  21.70 & 4.09 & 35.93  &            &           \\
            &     1 &  23.33 & 4.51 & 83.07  &            &           \\
            &     2 &  26.08 & 5.25 & 103.82 &            &           \\
            & Total &  24.05 & 6.90 & 225.40 &   25.6     &   392.45  \\
$^{197}$Au  &     0 &  22.43 & 4.15 & 49.61  &            &           \\
            &     1 &  22.86 & 4.26 & 103.06 &            &           \\
            &     2 &  23.81 & 4.50 & 111.83 &            &           \\
            & Total &  23.15 & 4.66 & 272.18 &   25.7     &   30.70   \\
\hline
\end{tabular}
\end{table}

\subsubsection{Estimation of the electric quadrupole moment}

We also estimated the intrinsic electric quadrupole moments $Q_0$ for the $^{165}$Ho and $^{169}$Tm ground states based on the present photoabsorption excitation functions and following the procedure detailed in Ref.~\cite{gheorghe_2026}. We expressed $Q_0$ in terms of the $d$ nuclear deformation parameter and the $R$ mean charge radius as
\begin{equation}
Q_0 =  \cfrac{2Z}{5} \Big( \cfrac{3R}{d+2} \Big)^2  (d^2-1). 
\end{equation}
Using the averaged best fit centroid energies of the first and second GDR peaks given in Table~\ref{table_GDR_par}, the hydrodynamic model predictions for the nuclear deformation parameter are $d$=1.305(15) for $^{165}$Ho and 1.310(13) for $^{169}$Tm. We computed the $R$=6.716(40)~fm and 6.746(5)~fm mean charge radii for $^{165}$Ho and $^{169}$Tm, respectively, from the experimentally determined root mean square $\sqrt{<r^2>}$ values of 5.202(31)~fm and 5.226(4)~fm respectively given in the Angeli and Marinova compilation~\cite{Angeli_2013}. 

We therefore extracted the $Q_0$=+7.00(34)~b value for the $^{165}$Ho ground state intrinsic electric quadrupole moment, and the +7.38(28)~b value for $^{169}$Tm. The positive $Q_0$ signs are deduced from the $R_A$ ratio of the area under the upper energy component of the GDR to that under the lower energy component, where the value of 2 is predicted for prolate spheroidal nuclei and 0.5 for oblate ones~\cite{danos_1958}. Present $R_A$ experimental values of 1.44 and 2.28 were obtained for $^{165}$Ho and $^{169}$Tm, respectively, based on the SLO fit to the total photoabsorption cross section. We further related the spectroscopic quadrupole electric moment to the intrinsic one~\cite{bohr_mottelson} and obtained the $Q_I$=+3.27(16)~b value for the $J^\Pi=7/2^-$ ground state of $^{165}$Ho from the present $Q_0$. This is lower than the $+3.58(2)\,\mathrm{b}$ value~\cite{Olaniyi_1982} obtained by measuring X rays in pionic holmium atoms, which is the value recommended by the INDC compilation from Ref.~\cite{stone_2021}. The $J^\Pi=1/2^+$ ground state of $^{169}$Tm has no spectroscopic electric quadrupole moment as $J = 1/2$ forbids it. 

%https://www-nds.iaea.org/nuclearmoments/isotope_measurement_results.php?A=169&Z=69

\subsection{Statistical model calculations for photoneutron reactions}
\label{sec:empire}

The new experimental data are now compared with statistical model calculations obtained with the EMPIRE~3.2.3 and the TALYS reaction codes. For both cross sections and average energy results, we extracted the contributions corresponding to the reactions with emission of $i$ neutrons regardless of the existence of accompanying charged particles, as described in Sec.~V.C of Ref.~\cite{kawano_2020}. 

\begin{figure*}[t]
\centering
\includegraphics[width=0.92\textwidth, angle=0]{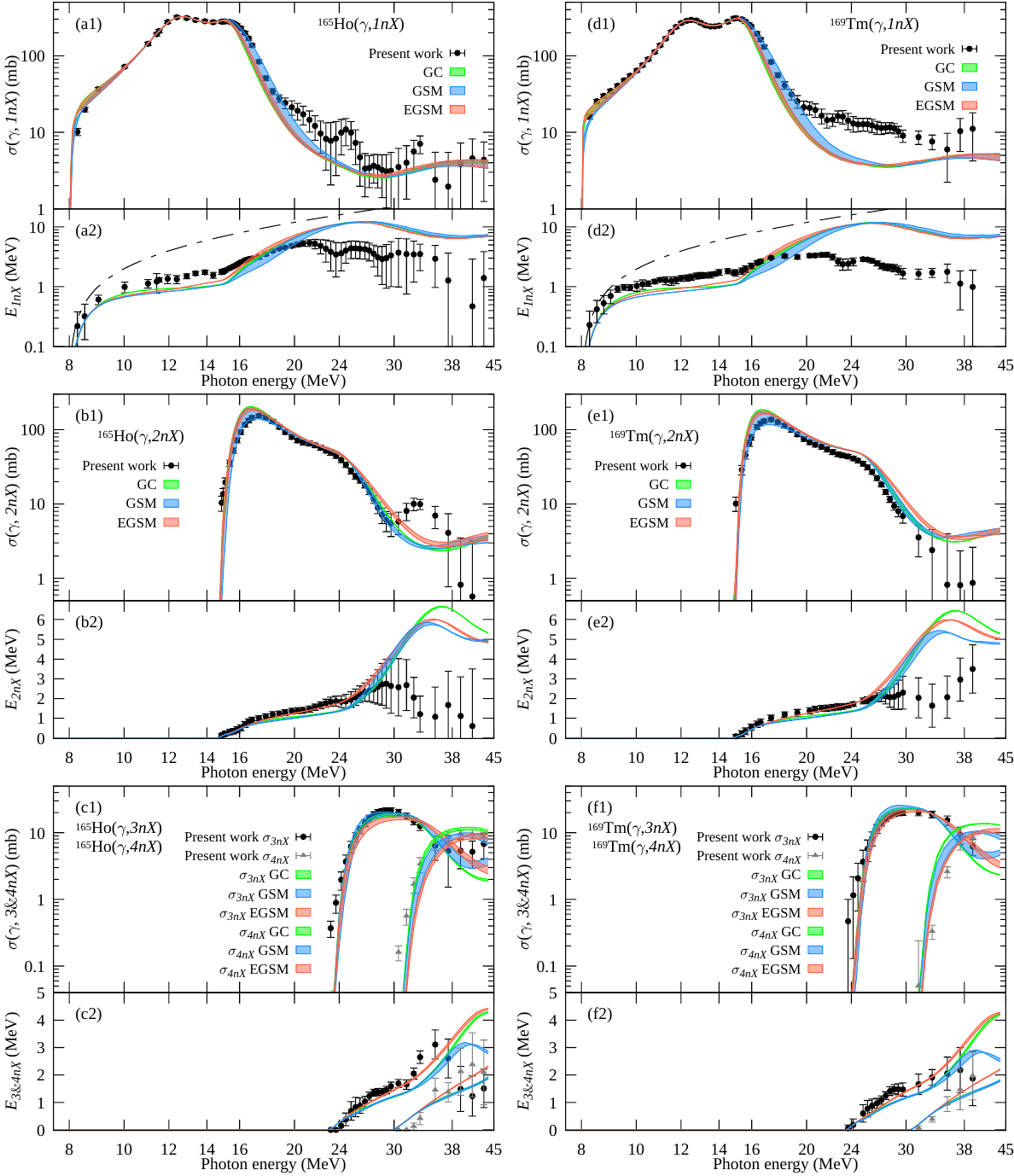}
\caption{Present $(\gamma,\,inX)$ experimental data compared with EMPIRE statistical model calculations performed with the GC, EGSM and GSM models of NLDs. The shaded areas correspond to the sensitivity to the PSF model. The dash-dotted lines in (a2) and (d2) correspond to the maximum $(\gamma,\,n)$ neutron energies given by kinematics (A-1)/A$\cdot$($E_\gamma-S_n$). Color online.} \label{fig_cs_en_gxn_band}
\end{figure*}

\subsubsection{EMPIRE sensitivity test to phenomenological NLDs and PSF parametrizations}

In our recent photoneutron study on $^{159}$Tb~\cite{gheorghe_2026}, we noticed that the choice in the PSF model affected not only the total photoabsorption cross sections, but also the competition between the partial $(\gamma,\,inX)$ reactions, which is expected to be mainly determined by the nuclear level densities~\cite{kawano_2020}.  
Thus, we now tested the sensitivity of the $^{165}$Ho and $^{169}$Tm $(\gamma,\,inX)$ cross sections and average neutron energies to the combinations of different phenomenological nuclear level densities and photon strength functions available in the EMPIRE code. After testing several optical model potentials for neutrons, we adopted the Koning-Delaroche general spherical one (RIPL ID - 2405). The PCROSS model~\cite{capote_1991_pcross} was used to compute the first neutron preequilibrium contribution.

Figure~\ref{fig_cs_en_gxn_band} shows the present $(\gamma,\,inX)$ experimental data (full circles) compared with EMPIRE statistical model calculations performed with the following level density approaches:
\begin{itemize}
\item enhanced generalized superfluid (EGSM) model~\cite{herman_2007_empire}, representing the EMPIRE-specific NLDs;
\item generalized superfluid (GSM) model; 
\item Gilbert-Cameron (GC) model. 
\end{itemize}
Calculations were performed with each of the above NLDs in combination with the three SLO, MLO2 and SMLO PSF models tested in the previous section. 
We mention that the EMPIRE code uses the same PSF model for all the compound and residual nuclei involved in the modeled reactions. For the target nuclei $^{165}$Ho and $^{169}$Tm we used the experimentally constrained PSF parameters listed in Table~\ref{table_GDR_par}, while the PSF parameterizations for the remaining nuclei were retrieved from RIPL~\cite{capote2009_ripl}. 

The hashed areas in Fig.~\ref{fig_cs_en_gxn_band} corresponding to each of the NLDs show the sensitivity to the PSF model. We note that the competition between the $(\gamma,\,inX)$ channels is influenced by both the choice of the NLD and of the PSF, where the results most sensitive to the photon strength function model are those of the calculations performed with the GSM level density, both for cross sections and for average energies.

We notice that the results obtained with the same NLD model are consistent for the two nuclei investigated. For both $^{165}$Ho and $^{169}$Tm, the GSM level density best describes the competition between the $(\gamma,\,1n)$ and $(\gamma,\,2n)$ channels at excitation energies immediately above the $S_{2n}$, while the GC and GSM densities overestimate the $(\gamma,\,2n)$ channel at the peak region. The GSM calculations also reproduce well the high-energy tail of the $(\gamma,\,2nX)$ cross section, especially for $^{165}$Ho, as well as the $(\gamma,\,3nX)$ and $(\gamma,\,4nX)$ cross sections over the entire investigated energy range. The GC density calculations describe fairly well the high-energy region of the $(\gamma,2nX)$ cross section, but do not reproduce the competition between the $(\gamma,\,3nX)$ and $(\gamma,\,4nX)$ ones. The EGSM calculations overestimate the $(\gamma,\,2nX)$ cross section at incident energies greater than 24 MeV and underestimate the $(\gamma,\,3nX)$ and $(\gamma,\,4nX)$ ones. A difference between the calculations for the two nuclei is observed at the high-energy tail of the $(\gamma,\,1nX)$ cross section, which is well described by all calculations for $^{165}$Ho and underestimated in all cases for $^{169}$Tm.
 
Next, we compare the experimental average photoneutron energies with the results of the EMPIRE statistical model calculations. The $(\gamma,\,1nX)$ and $(\gamma,\,2nX)$ reactions were investigated over a sufficiently wide energy range to observe that both the experimental and calculated average photoneutron energies present an increasing region characterized by two slopes, first a smooth one that becomes steeper after the $(\gamma,\,2n)$ and $(\gamma,\,3n)$ thresholds, respectively, reaching a maximum followed by a smooth decrease. However, although the calculations describe the experimental data qualitatively well, quantitatively they describe satisfactorily only the first increasing slope and the beginning of the second one. The maxima of approximately 12~MeV predicted by the statistical model calculations for the $E_{1nX}$ energies are significantly higher than those recorded experimentally for both $^{165}$Ho, of 5.4~MeV, and for $^{169}$Tm, of 3.4~MeV. Similarly, calculations predict a maximum average energy of approximately 6~MeV for neutrons emitted in the $(\gamma,\,2nX)$ reaction, while experimentally we recorded maxima of the order of 3~MeV for both nuclei. Although the results obtained with the three NLD inputs show strong common features, there are also important differences in the modeling of neutron energies in the increasing, low excitation energy region. We observe that EGSM describes well the average energies of neutrons emitted in the $(\gamma,\,2n)$, $(\gamma,\,3n)$ and $(\gamma,\,4n)$ reactions in the low excitation region, while those with GSM and GC systematically underestimate them. 

The high kinetic energy of $^{169}$Tm$(\gamma,\,n)$ neutrons in the vicinity of the $S_n$ is not reproduced by any calculations. A possible source of such non-statistical neutrons could have been a direct neutron knockout reaction (DKO). We used the model of direct surface nuclear photoeffect on heavy deformed nuclei \cite{Orlin_2023} to estimate the contribution of such processes. However, the calculated maximum cross section of neutron DKO on $^{169}$Tm was only about 2.5 mb at 13 MeV. Therefore, since the DKO of near-surface nucleons with excitation of rotational motion comprises the major part of direct photoneutron reaction cross section, the contribution of direct neutron emission can be considered negligible.

In summary, EMPIRE calculations describe best the present $^{165}$Ho and $^{169}$Tm photoneutron cross sections when using the GSM nuclear level density model. However, the corresponding experimental photoneutron average energies are best described using the EGSM model. As the calculation results are also fairly sensitive to the PSF model, we identified that the combination of GSM+MLO2 describes best the cross sections, while the EGSM+SLO best describes the average neutron energies. The resulting calculations performed with the two combinations are represented alongside present and existing experimental data in Fig.~\ref{fig_cs_en_gxn}. We note that a simultaneous improvement in the description of the $(\gamma,\,inX)$ cross sections and average neutron energies could not be obtained by varying the optical model potentials and the NLDs and preequilibrium parameters.

\subsubsection{TALYS calculations with microscopic NLDs}

\begin{figure*}[t]
\centering
\includegraphics[width=0.92\textwidth, angle=0]{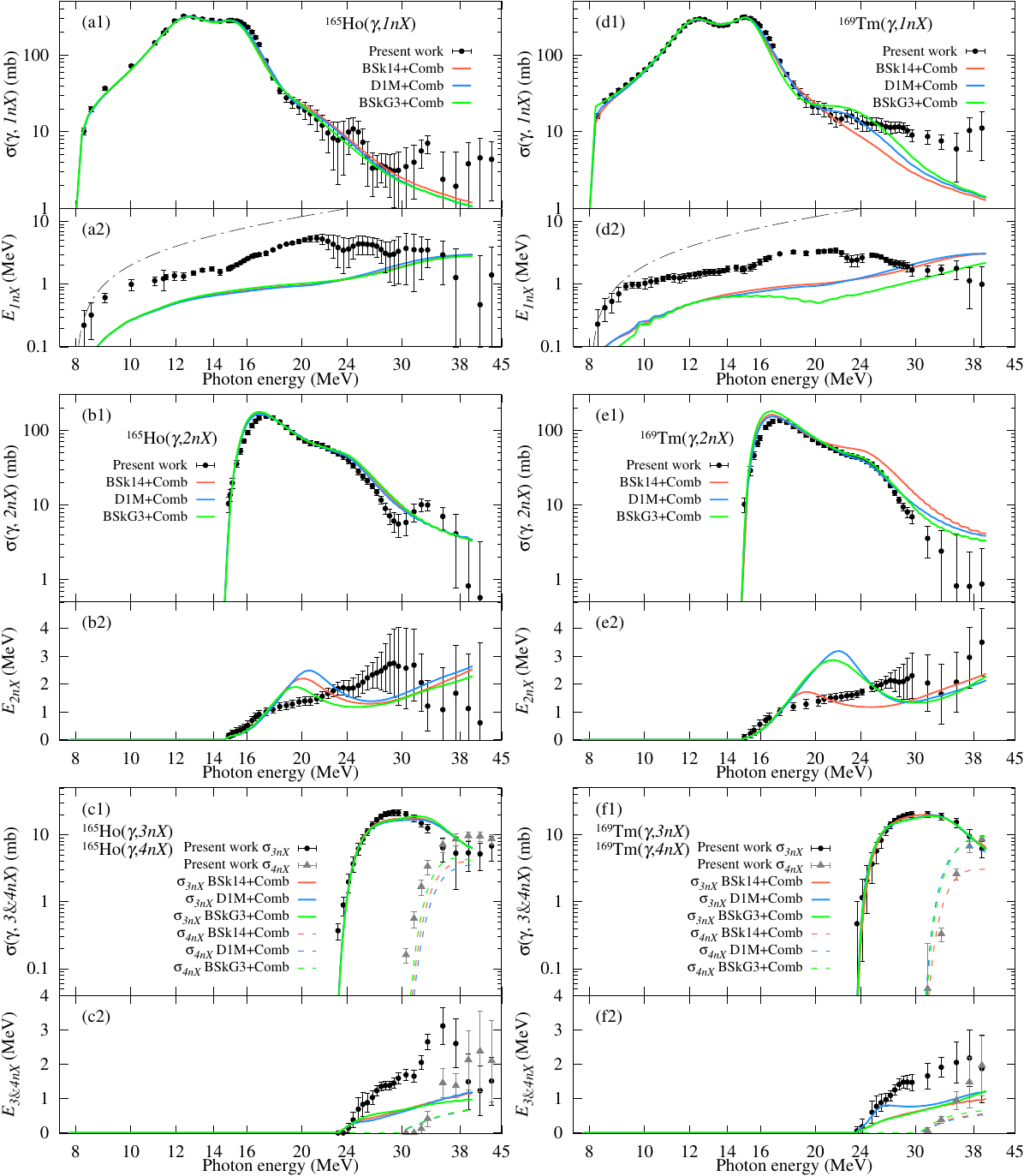}
\caption{Present $(\gamma,\,inX)$ experimental data compared with TALYS statistical model calculations using microscopic nuclear level density models, namely BSk14+Comb~\cite{goriely_2008}, D1M+Comb~\cite{hilaire_2012} and BSkG3+Comb~\cite{goriely_2026}. The dash-dotted lines in (a2) and (d2) correspond to the maximum $(\gamma,\,n)$ neutron energies given by kinematics (A-1)/A$\cdot$($E_\gamma-S_n$). Color online.} \label{fig_cs_en_gxn_Talys}
\end{figure*}

Calculations similar to the EMPIRE ones described in Sec.~\ref{sec:empire} were performed with the TALYS reaction code \cite{koning_2023_talys} adopting the microscopic nuclear density models, namely the BSk14 plus combinatorial model \cite{goriely_2008}, the combinatorial model based on a temperature-dependent Hartree-Fock-Bogolyubov (HFB) calculation and the Gogny D1M interaction \cite{hilaire_2012} and the recent triaxial HFB plus combinatorial model \cite{goriely_2026} with the BSkG3 energy density functional. These models are all based on the combinatorial approach but may differ significantly in the associated ground state properties (mainly single-particle level scheme, pairing properties and deformation) on which the nuclear level density is built. The nuclear level densities in the residual Ho and Tm isotopes were adjusted to optimise the description of the partial $(\gamma,inX)$ cross sections using the $\alpha$ and $\delta$ renormalisation parameters affecting the nuclear entropy and pairing, respectively, as detailed in Ref.~\cite{goriely_2026}. Since the various PSF variants fitted to the total photoabsorption cross section were found in Sec.~\ref{sec:empire} not to affect significantly the partial $(\gamma,inX)$ cross sections, the MLO2 functional was adopted in the TALYS calculations. 

As shown in Fig.~\ref{fig_cs_en_gxn_Talys}, the TALYS descriptions of the partial $(\gamma,inX)$ cross sections present relatively similar patterns as those found with EMPIRE calculations. In particular the $(\gamma,1nX)$ and $(\gamma,2nX)$ cross sections tend to be underestimated and overestimated, respectively, above typically some 20~MeV. Using different nuclear level density combinatorial models essentially gives rise to similar descriptions, except for the $^{169}$Tm $(\gamma,2nX)$ cross section which is quite strongly overestimated with the BSk14 plus combinatorial model \cite{goriely_2008}. While the $^{169}$Tm $(\gamma,3nX)$ and $(\gamma,4nX)$ cross sections are rather well reproduced, except for the underestimate of the $4n$ channel with the BSk14 plus combinatorial model, more discrepancies are found for $^{165}$Ho cross sections, especially for the $4n$ channel, irrespective of the nuclear level density model.

As far as the average photoneutron energies are concerned, TALYS tends to systematically underestimate experimental data, especially in the vicinity of the threshold energies, except for the $2n$ channel for which  TALYS overestimates data around 20~MeV (while EMPIRE overestimates data above 30 MeV). The EMPIRE description of the photoneutron energies is found to be in much better agreement with data in comparison with TALYS. The origin of these discrepancies may not come from the nuclear level densities, since as shown in Fig.~\ref{fig_cs_en_gxn_Talys}, the different level density models  give rise to essentially similar global behaviours.

\section{Summary and conclusions} \label{sec_summary} 

Photoneutron reactions were investigated in the giant dipole resonance region for the deformed $^{165}$Ho and $^{169}$Tm nuclei using quasimonochromatic laser Compton scattering $\gamma$-ray beams at the NewSUBARU facility. We obtained $(\gamma,\,1nX)$, $(\gamma,\,2nX)$, $(\gamma,\,3nX)$ and $(\gamma,\,4nX)$ reactions cross sections and average neutron emission energies by the use of a high-and-flat efficiency moderated array of $^3$He counters and an associated neutron multiplicity sorting method. 

The present $^{165}$Ho results were compared to existing data. Although the strength of the present results is in good agreement with the Saclay data of Refs.~\cite{Bergere_1968,Axel_1966}, we notice a discrepancy in the neutron multiplicity sorting procedures, which suggests that the Saclay group may have allocated $(\gamma,\,2nX)$ events to the $(\gamma,\,1nX)$ channel. We also observed an energy shift of $\sim$0.35~MeV between our results and the Saclay ones. Besides a uniform $\sim$10$\%$ strength difference, the present $(\gamma,\,inX)$ cross sections and the Livermore ones of Ref.~\cite{Berman_1969} are in good agreement on the entire investigated energy range.

The total photoneutron cross sections $\sigma(\gamma,\,Sn)$ were determined by summing the partial $(\gamma,\,inX)$ components. GDR parameters based on phenomenological Lorentzian models were extracted by fitting the present $\sigma(\gamma,\,Sn)$ data with adjustments for the missing contribution of charged-particle-only reactions not observed experimentally. For both nuclei we observed high energy structures at $\sim$20-25~MeV which correspond to the giant quadrupole resonance predictions of the microscopic KMFR model. From the present and the recent $^{197}$Au~\cite{gheorghe_2025} and $^{159}$Tb~\cite{gheorghe_2026} NewSUBARU results, we notice that the experimental results don't follow the predicted GQR centroid energy behavior of a decrease with the increase in the mass number $A$. Based on the present centroid energies of the first and second GDR peaks, we obtained the hydrodynamic model predictions of +7.00(34)~b and +7.38(28)~b for the intrinsic electric quadrupole moments of the ground states of $^{165}$Ho and $^{169}$Tm, respectively. 

The present experimental excitation functions and photoneutron energies were compared to statistical model calculations.
Using the EMPIRE statistical model code, we performed a sensitivity test to phenomenological models of photon strength functions and nuclear level densities. For both nuclei, the EMPIRE calculations reproduce best the partial $(\gamma,\,inX)$ cross sections using the combination of the GSM and MLO2 models for the NLDs and PSF respectively, while the average photoneutron energies are best described with the combination of the EGSM and SLO models. The TALYS code was used for reproducing the present experimental $(\gamma,\,inX)$ cross sections and average neutron energies using microscopic nuclear level density models. While the cross sections are rather well described, the photoneutron energies are quite systematically underestimated. 

All the experimental results obtained in the present paper are available in numerical format in the Supplemental Material~\cite{supplemental_material}.

\section{Acknowledgments}
The authors are grateful to H. Ohgaki of the Institute of Advanced Energy, Kyoto University for making a large volume LaBr$_3$:Ce detector available for the experiment.  
We sincerely thank J.E.~Midtb\o{} for helping with the experiment. 
I.G., D.F. and M.K. acknowledge the support from the Extreme Light Infrastructure Nuclear Physics (ELI-NP) Phase II, a project cofinanced by the Romanian Government and the European Union through the European Regional Development Fund - the Competitiveness Operational Programme (1/07.07.2016, COP, ID 1334).
I.G. and D.F. acknowledge the support from the Romanian Project PN-23-21-01-02.
S.G. acknowledges financial support from F.R.S.-FNRS (Belgium). This work was supported by the F.R.S.- FNRS and FWO under the EOS Project Nr O022818F. 
This work has been funded in part by the Research Council of Norway through its grant to the Norwegian Nuclear Research Centre (Project No. 341985) and through Project No. 325714.

\bibliography{references}{}

\providecommand{\noopsort}[1]{}\providecommand{\singleletter}[1]{#1}%
\begin{thebibliography}{59}%
\makeatletter
\providecommand \@ifxundefined [1]{%
 \@ifx{#1\undefined}
}%
\providecommand \@ifnum [1]{%
 \ifnum #1\expandafter \@firstoftwo
 \else \expandafter \@secondoftwo
 \fi
}%
\providecommand \@ifx [1]{%
 \ifx #1\expandafter \@firstoftwo
 \else \expandafter \@secondoftwo
 \fi
}%
\providecommand \natexlab [1]{#1}%
\providecommand \enquote  [1]{``#1''}%
\providecommand \bibnamefont  [1]{#1}%
\providecommand \bibfnamefont [1]{#1}%
\providecommand \citenamefont [1]{#1}%
\providecommand \href@noop [0]{\@secondoftwo}%
\providecommand \href [0]{\begingroup \@sanitize@url \@href}%
\providecommand \@href[1]{\@@startlink{#1}\@@href}%
\providecommand \@@href[1]{\endgroup#1\@@endlink}%
\providecommand \@sanitize@url [0]{\catcode `\\12\catcode `\$12\catcode
  `\&12\catcode `\#12\catcode `\^12\catcode `\_12\catcode `\%12\relax}%
\providecommand \@@startlink[1]{}%
\providecommand \@@endlink[0]{}%
\providecommand \url  [0]{\begingroup\@sanitize@url \@url }%
\providecommand \@url [1]{\endgroup\@href {#1}{\urlprefix }}%
\providecommand \urlprefix  [0]{URL }%
\providecommand \Eprint [0]{\href }%
\providecommand \doibase [0]{https://doi.org/}%
\providecommand \selectlanguage [0]{\@gobble}%
\providecommand \bibinfo  [0]{\@secondoftwo}%
\providecommand \bibfield  [0]{\@secondoftwo}%
\providecommand \translation [1]{[#1]}%
\providecommand \BibitemOpen [0]{}%
\providecommand \bibitemStop [0]{}%
\providecommand \bibitemNoStop [0]{.\EOS\space}%
\providecommand \EOS [0]{\spacefactor3000\relax}%
\providecommand \BibitemShut  [1]{\csname bibitem#1\endcsname}%
\let\auto@bib@innerbib\@empty
%</preamble>
\bibitem [{\citenamefont {Gheorghe}\ \emph {et~al.}(2017)\citenamefont
  {Gheorghe}, \citenamefont {Utsunomiya}, \citenamefont {Katayama},
  \citenamefont {Filipescu}, \citenamefont {Belyshev}, \citenamefont {Stopani},
  \citenamefont {Orlin}, \citenamefont {Varlamov}, \citenamefont {Shima},
  \citenamefont {Amano}, \citenamefont {Miyamoto}, \citenamefont {Lui},
  \citenamefont {Kawano},\ and\ \citenamefont {Goriely}}]{gheorghe_2017}%
  \BibitemOpen
  \bibfield  {author} {\bibinfo {author} {\bibfnamefont {I.}~\bibnamefont
  {Gheorghe}}, \bibinfo {author} {\bibfnamefont {H.}~\bibnamefont
  {Utsunomiya}}, \bibinfo {author} {\bibfnamefont {S.}~\bibnamefont
  {Katayama}}, \bibinfo {author} {\bibfnamefont {D.}~\bibnamefont {Filipescu}},
  \bibinfo {author} {\bibfnamefont {S.}~\bibnamefont {Belyshev}}, \bibinfo
  {author} {\bibfnamefont {K.}~\bibnamefont {Stopani}}, \bibinfo {author}
  {\bibfnamefont {V.}~\bibnamefont {Orlin}}, \bibinfo {author} {\bibfnamefont
  {V.}~\bibnamefont {Varlamov}}, \bibinfo {author} {\bibfnamefont
  {T.}~\bibnamefont {Shima}}, \bibinfo {author} {\bibfnamefont
  {S.}~\bibnamefont {Amano}}, \bibinfo {author} {\bibfnamefont
  {S.}~\bibnamefont {Miyamoto}}, \bibinfo {author} {\bibfnamefont {Y.-W.}\
  \bibnamefont {Lui}}, \bibinfo {author} {\bibfnamefont {T.}~\bibnamefont
  {Kawano}},\ and\ \bibinfo {author} {\bibfnamefont {S.}~\bibnamefont
  {Goriely}},\ }\bibfield  {title} {\bibinfo {title} {Photoneutron
  cross-section measurements in the
  $^{209}\mathrm{Bi}$($\ensuremath{\gamma},xn$) reaction with a new method of
  direct neutron-multiplicity sorting},\ }\href
  {https://doi.org/10.1103/PhysRevC.96.044604} {\bibfield  {journal} {\bibinfo
  {journal} {Phys. Rev. C}\ }\textbf {\bibinfo {volume} {96}},\ \bibinfo
  {pages} {044604} (\bibinfo {year} {2017})},\ \bibinfo {note} {erratum:
  \textit{Phys. Rev. C} \textbf{99}, 059901(E) (2019),
  \url{https://doi.org/10.1103/PhysRevC.99.059901}}\BibitemShut {NoStop}%
\bibitem [{\citenamefont {Filipescu}\ \emph {et~al.}(2024)\citenamefont
  {Filipescu}, \citenamefont {Gheorghe}, \citenamefont {Goriely}, \citenamefont
  {Tudora}, \citenamefont {Nishio}, \citenamefont {Ohtsuki}, \citenamefont
  {Wang}, \citenamefont {Fan}, \citenamefont {Stopani}, \citenamefont {Suzaki},
  \citenamefont {Hirose}, \citenamefont {Inagaki}, \citenamefont {Lui},
  \citenamefont {Ari-izumi}, \citenamefont {Miyamoto}, \citenamefont {Otsuka},\
  and\ \citenamefont {Utsunomiya}}]{filipescu_2024}%
  \BibitemOpen
  \bibfield  {author} {\bibinfo {author} {\bibfnamefont {D.}~\bibnamefont
  {Filipescu}}, \bibinfo {author} {\bibfnamefont {I.}~\bibnamefont {Gheorghe}},
  \bibinfo {author} {\bibfnamefont {S.}~\bibnamefont {Goriely}}, \bibinfo
  {author} {\bibfnamefont {A.}~\bibnamefont {Tudora}}, \bibinfo {author}
  {\bibfnamefont {K.}~\bibnamefont {Nishio}}, \bibinfo {author} {\bibfnamefont
  {T.}~\bibnamefont {Ohtsuki}}, \bibinfo {author} {\bibfnamefont
  {H.}~\bibnamefont {Wang}}, \bibinfo {author} {\bibfnamefont {G.}~\bibnamefont
  {Fan}}, \bibinfo {author} {\bibfnamefont {K.}~\bibnamefont {Stopani}},
  \bibinfo {author} {\bibfnamefont {F.}~\bibnamefont {Suzaki}}, \bibinfo
  {author} {\bibfnamefont {K.}~\bibnamefont {Hirose}}, \bibinfo {author}
  {\bibfnamefont {M.}~\bibnamefont {Inagaki}}, \bibinfo {author} {\bibfnamefont
  {Y.-W.}\ \bibnamefont {Lui}}, \bibinfo {author} {\bibfnamefont
  {T.}~\bibnamefont {Ari-izumi}}, \bibinfo {author} {\bibfnamefont
  {S.}~\bibnamefont {Miyamoto}}, \bibinfo {author} {\bibfnamefont
  {T.}~\bibnamefont {Otsuka}},\ and\ \bibinfo {author} {\bibfnamefont
  {H.}~\bibnamefont {Utsunomiya}},\ }\bibfield  {title} {\bibinfo {title}
  {Giant dipole resonance photofission and photoneutron reactions in
  $^{238}\mathrm{U}$ and $^{232}\mathrm{Th}$},\ }\href
  {https://doi.org/10.1103/PhysRevC.109.044602} {\bibfield  {journal} {\bibinfo
   {journal} {Phys. Rev. C}\ }\textbf {\bibinfo {volume} {109}},\ \bibinfo
  {pages} {044602} (\bibinfo {year} {2024})}\BibitemShut {NoStop}%
\bibitem [{\citenamefont {Gheorghe}\ \emph {et~al.}(2024)\citenamefont
  {Gheorghe}, \citenamefont {Goriely}, \citenamefont {Wagner}, \citenamefont
  {Aumann}, \citenamefont {Baumann}, \citenamefont {van Beek}, \citenamefont
  {Kuchenbrod}, \citenamefont {Scheit}, \citenamefont {Symochko}, \citenamefont
  {Ari-izumi}, \citenamefont {Bello~Garrote}, \citenamefont {Eriksen},
  \citenamefont {Paulsen}, \citenamefont {Pedersen}, \citenamefont {Reaz},
  \citenamefont {Ingeberg}, \citenamefont {Belyshev}, \citenamefont
  {Miyamoto},\ and\ \citenamefont {Utsunomiya}}]{gheorghe_2024}%
  \BibitemOpen
  \bibfield  {author} {\bibinfo {author} {\bibfnamefont {I.}~\bibnamefont
  {Gheorghe}}, \bibinfo {author} {\bibfnamefont {S.}~\bibnamefont {Goriely}},
  \bibinfo {author} {\bibfnamefont {N.}~\bibnamefont {Wagner}}, \bibinfo
  {author} {\bibfnamefont {T.}~\bibnamefont {Aumann}}, \bibinfo {author}
  {\bibfnamefont {M.}~\bibnamefont {Baumann}}, \bibinfo {author} {\bibfnamefont
  {P.}~\bibnamefont {van Beek}}, \bibinfo {author} {\bibfnamefont
  {P.}~\bibnamefont {Kuchenbrod}}, \bibinfo {author} {\bibfnamefont
  {H.}~\bibnamefont {Scheit}}, \bibinfo {author} {\bibfnamefont
  {D.}~\bibnamefont {Symochko}}, \bibinfo {author} {\bibfnamefont
  {T.}~\bibnamefont {Ari-izumi}}, \bibinfo {author} {\bibfnamefont {F.~L.}\
  \bibnamefont {Bello~Garrote}}, \bibinfo {author} {\bibfnamefont
  {T.}~\bibnamefont {Eriksen}}, \bibinfo {author} {\bibfnamefont
  {W.}~\bibnamefont {Paulsen}}, \bibinfo {author} {\bibfnamefont {L.~G.}\
  \bibnamefont {Pedersen}}, \bibinfo {author} {\bibfnamefont {F.}~\bibnamefont
  {Reaz}}, \bibinfo {author} {\bibfnamefont {V.~W.}\ \bibnamefont {Ingeberg}},
  \bibinfo {author} {\bibfnamefont {S.}~\bibnamefont {Belyshev}}, \bibinfo
  {author} {\bibfnamefont {S.}~\bibnamefont {Miyamoto}},\ and\ \bibinfo
  {author} {\bibfnamefont {H.}~\bibnamefont {Utsunomiya}},\ }\bibfield  {title}
  {\bibinfo {title} {Photoneutron cross section measurements on
  $^{208}\mathrm{Pb}$ in the giant dipole resonance region},\ }\href
  {https://doi.org/10.1103/PhysRevC.110.014619} {\bibfield  {journal} {\bibinfo
   {journal} {Phys. Rev. C}\ }\textbf {\bibinfo {volume} {110}},\ \bibinfo
  {pages} {014619} (\bibinfo {year} {2024})}\BibitemShut {NoStop}%
\bibitem [{\citenamefont {Gheorghe}\ \emph {et~al.}(2025)\citenamefont
  {Gheorghe}, \citenamefont {Ari-izumi}, \citenamefont {Goriely}, \citenamefont
  {Filipescu}, \citenamefont {Belyshev}, \citenamefont {Stopani}, \citenamefont
  {Wang}, \citenamefont {Fan}, \citenamefont {Scheit}, \citenamefont
  {Symochko}, \citenamefont {Krzysiek}, \citenamefont {Renstr\o{}m},
  \citenamefont {Tveten}, \citenamefont {Miyamoto},\ and\ \citenamefont
  {Utsunomiya}}]{gheorghe_2025}%
  \BibitemOpen
  \bibfield  {author} {\bibinfo {author} {\bibfnamefont {I.}~\bibnamefont
  {Gheorghe}}, \bibinfo {author} {\bibfnamefont {T.}~\bibnamefont {Ari-izumi}},
  \bibinfo {author} {\bibfnamefont {S.}~\bibnamefont {Goriely}}, \bibinfo
  {author} {\bibfnamefont {D.}~\bibnamefont {Filipescu}}, \bibinfo {author}
  {\bibfnamefont {S.}~\bibnamefont {Belyshev}}, \bibinfo {author}
  {\bibfnamefont {K.}~\bibnamefont {Stopani}}, \bibinfo {author} {\bibfnamefont
  {H.}~\bibnamefont {Wang}}, \bibinfo {author} {\bibfnamefont {G.}~\bibnamefont
  {Fan}}, \bibinfo {author} {\bibfnamefont {H.}~\bibnamefont {Scheit}},
  \bibinfo {author} {\bibfnamefont {D.}~\bibnamefont {Symochko}}, \bibinfo
  {author} {\bibfnamefont {M.}~\bibnamefont {Krzysiek}}, \bibinfo {author}
  {\bibfnamefont {T.}~\bibnamefont {Renstr\o{}m}}, \bibinfo {author}
  {\bibfnamefont {G.~M.}\ \bibnamefont {Tveten}}, \bibinfo {author}
  {\bibfnamefont {S.}~\bibnamefont {Miyamoto}},\ and\ \bibinfo {author}
  {\bibfnamefont {H.}~\bibnamefont {Utsunomiya}},\ }\bibfield  {title}
  {\bibinfo {title} {Photoneutron reactions on gold in the giant dipole
  resonance region: Reaction cross sections and average kinetic energies of
  $(\ensuremath{\gamma},xn)$ photoneutrons},\ }\href
  {https://doi.org/10.1103/PhysRevC.111.014611} {\bibfield  {journal} {\bibinfo
   {journal} {Phys. Rev. C}\ }\textbf {\bibinfo {volume} {111}},\ \bibinfo
  {pages} {014611} (\bibinfo {year} {2025})}\BibitemShut {NoStop}%
\bibitem [{\citenamefont {Gheorghe}\ \emph {et~al.}(2026)\citenamefont
  {Gheorghe}, \citenamefont {Goriely}, \citenamefont {Stopani}, \citenamefont
  {Belyshev}, \citenamefont {Fan}, \citenamefont {Liu}, \citenamefont
  {Krzysiek}, \citenamefont {Filipescu}, \citenamefont {Ari-izumi},
  \citenamefont {Miyamoto},\ and\ \citenamefont {Utsunomiya}}]{gheorghe_2026}%
  \BibitemOpen
  \bibfield  {author} {\bibinfo {author} {\bibfnamefont {I.}~\bibnamefont
  {Gheorghe}}, \bibinfo {author} {\bibfnamefont {S.}~\bibnamefont {Goriely}},
  \bibinfo {author} {\bibfnamefont {K.}~\bibnamefont {Stopani}}, \bibinfo
  {author} {\bibfnamefont {S.}~\bibnamefont {Belyshev}}, \bibinfo {author}
  {\bibfnamefont {G.}~\bibnamefont {Fan}}, \bibinfo {author} {\bibfnamefont
  {L.}~\bibnamefont {Liu}}, \bibinfo {author} {\bibfnamefont {M.}~\bibnamefont
  {Krzysiek}}, \bibinfo {author} {\bibfnamefont {D.}~\bibnamefont {Filipescu}},
  \bibinfo {author} {\bibfnamefont {T.}~\bibnamefont {Ari-izumi}}, \bibinfo
  {author} {\bibfnamefont {S.}~\bibnamefont {Miyamoto}},\ and\ \bibinfo
  {author} {\bibfnamefont {H.}~\bibnamefont {Utsunomiya}},\ }\bibfield  {title}
  {\bibinfo {title} {Photoneutron reactions on $^{159}\mathrm{Tb}$ in the giant
  dipole resonance region: Reaction cross sections and average kinetic energies
  of ($\ensuremath{\gamma},xn$) photoneutrons},\ }\href
  {https://doi.org/10.1103/79bt-cxcv} {\bibfield  {journal} {\bibinfo
  {journal} {Phys. Rev. C}\ }\textbf {\bibinfo {volume} {113}},\ \bibinfo
  {pages} {044620} (\bibinfo {year} {2026})}\BibitemShut {NoStop}%
\bibitem [{\citenamefont {Amano}\ \emph {et~al.}(2009)\citenamefont {Amano},
  \citenamefont {Horikawa}, \citenamefont {Ishihara}, \citenamefont {Miyamoto},
  \citenamefont {Hayakawa}, \citenamefont {Shizuma},\ and\ \citenamefont
  {Mochizuki}}]{amano_2009}%
  \BibitemOpen
  \bibfield  {author} {\bibinfo {author} {\bibfnamefont {S.}~\bibnamefont
  {Amano}}, \bibinfo {author} {\bibfnamefont {K.}~\bibnamefont {Horikawa}},
  \bibinfo {author} {\bibfnamefont {K.}~\bibnamefont {Ishihara}}, \bibinfo
  {author} {\bibfnamefont {S.}~\bibnamefont {Miyamoto}}, \bibinfo {author}
  {\bibfnamefont {T.}~\bibnamefont {Hayakawa}}, \bibinfo {author}
  {\bibfnamefont {T.}~\bibnamefont {Shizuma}},\ and\ \bibinfo {author}
  {\bibfnamefont {T.}~\bibnamefont {Mochizuki}},\ }\bibfield  {title} {\bibinfo
  {title} {Several-{MeV} $\gamma$-ray generation at {NewSUBARU} by laser
  {Compton} backscattering},\ }\href
  {https://doi.org/https://doi.org/10.1016/j.nima.2009.01.010} {\bibfield
  {journal} {\bibinfo  {journal} {Nuclear Instruments and Methods in Physics
  Research Section A: Accelerators, Spectrometers, Detectors and Associated
  Equipment}\ }\textbf {\bibinfo {volume} {602}},\ \bibinfo {pages} {337}
  (\bibinfo {year} {2009})}\BibitemShut {NoStop}%
\bibitem [{\citenamefont {Horikawa}\ \emph {et~al.}(2010)\citenamefont
  {Horikawa}, \citenamefont {Miyamoto}, \citenamefont {Amano},\ and\
  \citenamefont {Mochizuki}}]{horikawa_2010}%
  \BibitemOpen
  \bibfield  {author} {\bibinfo {author} {\bibfnamefont {K.}~\bibnamefont
  {Horikawa}}, \bibinfo {author} {\bibfnamefont {S.}~\bibnamefont {Miyamoto}},
  \bibinfo {author} {\bibfnamefont {S.}~\bibnamefont {Amano}},\ and\ \bibinfo
  {author} {\bibfnamefont {T.}~\bibnamefont {Mochizuki}},\ }\bibfield  {title}
  {\bibinfo {title} {Measurements for the energy and flux of laser {Compton}
  scattering $\gamma$-ray photons generated in an electron storage ring:
  {NewSUBARU}},\ }\href
  {https://doi.org/https://doi.org/10.1016/j.nima.2010.02.259} {\bibfield
  {journal} {\bibinfo  {journal} {Nuclear Instruments and Methods in Physics
  Research Section A: Accelerators, Spectrometers, Detectors and Associated
  Equipment}\ }\textbf {\bibinfo {volume} {618}},\ \bibinfo {pages} {209}
  (\bibinfo {year} {2010})}\BibitemShut {NoStop}%
\bibitem [{\citenamefont {Kawano}\ \emph {et~al.}(2020)\citenamefont {Kawano},
  \citenamefont {Cho}, \citenamefont {Dimitriou}, \citenamefont {Filipescu},
  \citenamefont {Iwamoto}, \citenamefont {Plujko}, \citenamefont {Tao},
  \citenamefont {Utsunomiya}, \citenamefont {Varlamov}, \citenamefont {Xu},
  \citenamefont {Capote}, \citenamefont {Gheorghe}, \citenamefont
  {Gorbachenko}, \citenamefont {Jin}, \citenamefont {Renstr\o{}m},
  \citenamefont {Sin}, \citenamefont {Stopani}, \citenamefont {Tian},
  \citenamefont {Tveten}, \citenamefont {Wang}, \citenamefont {Belgya},
  \citenamefont {Firestone}, \citenamefont {Goriely}, \citenamefont {Kopecky},
  \citenamefont {Krti\v{c}ka}, \citenamefont {Schwengner}, \citenamefont
  {Siem},\ and\ \citenamefont {Wiedeking}}]{kawano_2020}%
  \BibitemOpen
  \bibfield  {author} {\bibinfo {author} {\bibfnamefont {T.}~\bibnamefont
  {Kawano}}, \bibinfo {author} {\bibfnamefont {Y.}~\bibnamefont {Cho}},
  \bibinfo {author} {\bibfnamefont {P.}~\bibnamefont {Dimitriou}}, \bibinfo
  {author} {\bibfnamefont {D.}~\bibnamefont {Filipescu}}, \bibinfo {author}
  {\bibfnamefont {N.}~\bibnamefont {Iwamoto}}, \bibinfo {author} {\bibfnamefont
  {V.}~\bibnamefont {Plujko}}, \bibinfo {author} {\bibfnamefont
  {X.}~\bibnamefont {Tao}}, \bibinfo {author} {\bibfnamefont {H.}~\bibnamefont
  {Utsunomiya}}, \bibinfo {author} {\bibfnamefont {V.}~\bibnamefont
  {Varlamov}}, \bibinfo {author} {\bibfnamefont {R.}~\bibnamefont {Xu}},
  \bibinfo {author} {\bibfnamefont {R.}~\bibnamefont {Capote}}, \bibinfo
  {author} {\bibfnamefont {I.}~\bibnamefont {Gheorghe}}, \bibinfo {author}
  {\bibfnamefont {O.}~\bibnamefont {Gorbachenko}}, \bibinfo {author}
  {\bibfnamefont {Y.}~\bibnamefont {Jin}}, \bibinfo {author} {\bibfnamefont
  {T.}~\bibnamefont {Renstr\o{}m}}, \bibinfo {author} {\bibfnamefont
  {M.}~\bibnamefont {Sin}}, \bibinfo {author} {\bibfnamefont {K.}~\bibnamefont
  {Stopani}}, \bibinfo {author} {\bibfnamefont {Y.}~\bibnamefont {Tian}},
  \bibinfo {author} {\bibfnamefont {G.}~\bibnamefont {Tveten}}, \bibinfo
  {author} {\bibfnamefont {J.}~\bibnamefont {Wang}}, \bibinfo {author}
  {\bibfnamefont {T.}~\bibnamefont {Belgya}}, \bibinfo {author} {\bibfnamefont
  {R.}~\bibnamefont {Firestone}}, \bibinfo {author} {\bibfnamefont
  {S.}~\bibnamefont {Goriely}}, \bibinfo {author} {\bibfnamefont
  {J.}~\bibnamefont {Kopecky}}, \bibinfo {author} {\bibfnamefont
  {M.}~\bibnamefont {Krti\v{c}ka}}, \bibinfo {author} {\bibfnamefont
  {R.}~\bibnamefont {Schwengner}}, \bibinfo {author} {\bibfnamefont
  {S.}~\bibnamefont {Siem}},\ and\ \bibinfo {author} {\bibfnamefont
  {M.}~\bibnamefont {Wiedeking}},\ }\bibfield  {title} {\bibinfo {title}
  {{IAEA} {Photonuclear} {Data} {Library} 2019},\ }\href
  {https://doi.org/https://doi.org/10.1016/j.nds.2019.12.002} {\bibfield
  {journal} {\bibinfo  {journal} {Nuclear Data Sheets}\ }\textbf {\bibinfo
  {volume} {163}},\ \bibinfo {pages} {109} (\bibinfo {year}
  {2020})}\BibitemShut {NoStop}%
\bibitem [{\citenamefont {Goriely}\ \emph {et~al.}(2019)\citenamefont
  {Goriely}, \citenamefont {Dimitriou}, \citenamefont {Wiedeking},
  \citenamefont {Belgya}, \citenamefont {Firestone}, \citenamefont {Kopecky},
  \citenamefont {Krti\v{c}ka}, \citenamefont {Plujko}, \citenamefont
  {Schwengner}, \citenamefont {Siem}, \citenamefont {Utsunomiya}, \citenamefont
  {Hilaire}, \citenamefont {P{\'e}ru}, \citenamefont {Cho}, \citenamefont
  {Filipescu}, \citenamefont {Iwamoto}, \citenamefont {Kawano}, \citenamefont
  {Varlamov},\ and\ \citenamefont {Xu}}]{goriely_2019}%
  \BibitemOpen
  \bibfield  {author} {\bibinfo {author} {\bibfnamefont {S.}~\bibnamefont
  {Goriely}}, \bibinfo {author} {\bibfnamefont {P.}~\bibnamefont {Dimitriou}},
  \bibinfo {author} {\bibfnamefont {M.}~\bibnamefont {Wiedeking}}, \bibinfo
  {author} {\bibfnamefont {T.}~\bibnamefont {Belgya}}, \bibinfo {author}
  {\bibfnamefont {R.}~\bibnamefont {Firestone}}, \bibinfo {author}
  {\bibfnamefont {J.}~\bibnamefont {Kopecky}}, \bibinfo {author} {\bibfnamefont
  {M.}~\bibnamefont {Krti\v{c}ka}}, \bibinfo {author} {\bibfnamefont
  {V.}~\bibnamefont {Plujko}}, \bibinfo {author} {\bibfnamefont
  {R.}~\bibnamefont {Schwengner}}, \bibinfo {author} {\bibfnamefont
  {S.}~\bibnamefont {Siem}}, \bibinfo {author} {\bibfnamefont {H.}~\bibnamefont
  {Utsunomiya}}, \bibinfo {author} {\bibfnamefont {S.}~\bibnamefont {Hilaire}},
  \bibinfo {author} {\bibfnamefont {S.}~\bibnamefont {P{\'e}ru}}, \bibinfo
  {author} {\bibfnamefont {Y.}~\bibnamefont {Cho}}, \bibinfo {author}
  {\bibfnamefont {D.}~\bibnamefont {Filipescu}}, \bibinfo {author}
  {\bibfnamefont {N.}~\bibnamefont {Iwamoto}}, \bibinfo {author} {\bibfnamefont
  {T.}~\bibnamefont {Kawano}}, \bibinfo {author} {\bibfnamefont
  {V.}~\bibnamefont {Varlamov}},\ and\ \bibinfo {author} {\bibfnamefont
  {R.}~\bibnamefont {Xu}},\ }\bibfield  {title} {\bibinfo {title} {Reference
  database for photon strength functions},\ }\href
  {https://doi.org/https://doi.org/10.1140/epja/i2019-12840-1} {\bibfield
  {journal} {\bibinfo  {journal} {The European Physical Journal A}\ }\textbf
  {\bibinfo {volume} {55}},\ \bibinfo {pages} {172} (\bibinfo {year}
  {2019})}\BibitemShut {NoStop}%
\bibitem [{\citenamefont {Goriely}\ \emph {et~al.}(2020)\citenamefont
  {Goriely}, \citenamefont {P\'eru}, \citenamefont {Col\`o}, \citenamefont
  {Roca-Maza}, \citenamefont {Gheorghe}, \citenamefont {Filipescu},\ and\
  \citenamefont {Utsunomiya}}]{goriely_2020}%
  \BibitemOpen
  \bibfield  {author} {\bibinfo {author} {\bibfnamefont {S.}~\bibnamefont
  {Goriely}}, \bibinfo {author} {\bibfnamefont {S.}~\bibnamefont {P\'eru}},
  \bibinfo {author} {\bibfnamefont {G.}~\bibnamefont {Col\`o}}, \bibinfo
  {author} {\bibfnamefont {X.}~\bibnamefont {Roca-Maza}}, \bibinfo {author}
  {\bibfnamefont {I.}~\bibnamefont {Gheorghe}}, \bibinfo {author}
  {\bibfnamefont {D.}~\bibnamefont {Filipescu}},\ and\ \bibinfo {author}
  {\bibfnamefont {H.}~\bibnamefont {Utsunomiya}},\ }\bibfield  {title}
  {\bibinfo {title} {{E1} moments from a coherent set of measured photoneutron
  cross sections},\ }\href {https://doi.org/10.1103/PhysRevC.102.064309}
  {\bibfield  {journal} {\bibinfo  {journal} {Phys. Rev. C}\ }\textbf {\bibinfo
  {volume} {102}},\ \bibinfo {pages} {064309} (\bibinfo {year}
  {2020})}\BibitemShut {NoStop}%
\bibitem [{\citenamefont {Danos}(1958)}]{danos_1958}%
  \BibitemOpen
  \bibfield  {author} {\bibinfo {author} {\bibfnamefont {M.}~\bibnamefont
  {Danos}},\ }\bibfield  {title} {\bibinfo {title} {On the long-range
  correlation model of the photonuclear effect},\ }\href
  {https://doi.org/https://doi.org/10.1016/0029-5582(58)90005-1} {\bibfield
  {journal} {\bibinfo  {journal} {Nuclear Physics}\ }\textbf {\bibinfo {volume}
  {5}},\ \bibinfo {pages} {23} (\bibinfo {year} {1958})}\BibitemShut {NoStop}%
\bibitem [{\citenamefont {Otsuka}\ \emph {et~al.}(2025)\citenamefont {Otsuka},
  \citenamefont {Tsunoda}, \citenamefont {Shimizu}, \citenamefont {Utsuno},
  \citenamefont {Abe},\ and\ \citenamefont {Ueno}}]{Otsuka_2025}%
  \BibitemOpen
  \bibfield  {author} {\bibinfo {author} {\bibfnamefont {T.}~\bibnamefont
  {Otsuka}}, \bibinfo {author} {\bibfnamefont {Y.}~\bibnamefont {Tsunoda}},
  \bibinfo {author} {\bibfnamefont {N.}~\bibnamefont {Shimizu}}, \bibinfo
  {author} {\bibfnamefont {Y.}~\bibnamefont {Utsuno}}, \bibinfo {author}
  {\bibfnamefont {T.}~\bibnamefont {Abe}},\ and\ \bibinfo {author}
  {\bibfnamefont {H.}~\bibnamefont {Ueno}},\ }\bibfield  {title} {\bibinfo
  {title} {Prevailing triaxial shapes in atomic nuclei and a quantum theory of
  rotation of composite objects},\ }\href
  {https://doi.org/10.1140/epja/s10050-025-01553-1} {\bibfield  {journal}
  {\bibinfo  {journal} {The European Physical Journal A}\ }\textbf {\bibinfo
  {volume} {61}},\ \bibinfo {pages} {126} (\bibinfo {year} {2025})}\BibitemShut
  {NoStop}%
\bibitem [{\citenamefont {Arnould}\ and\ \citenamefont
  {Goriely}(2003)}]{arnould_2003}%
  \BibitemOpen
  \bibfield  {author} {\bibinfo {author} {\bibfnamefont {M.}~\bibnamefont
  {Arnould}}\ and\ \bibinfo {author} {\bibfnamefont {S.}~\bibnamefont
  {Goriely}},\ }\bibfield  {title} {\bibinfo {title} {The p-process of stellar
  nucleosynthesis: astrophysics and nuclear physics status},\ }\href
  {https://doi.org/https://doi.org/10.1016/S0370-1573(03)00242-4} {\bibfield
  {journal} {\bibinfo  {journal} {Physics Reports}\ }\textbf {\bibinfo {volume}
  {384}},\ \bibinfo {pages} {1} (\bibinfo {year} {2003})}\BibitemShut {NoStop}%
\bibitem [{\citenamefont {Arnould}\ and\ \citenamefont
  {Goriely}(2020)}]{arnould_2020}%
  \BibitemOpen
  \bibfield  {author} {\bibinfo {author} {\bibfnamefont {M.}~\bibnamefont
  {Arnould}}\ and\ \bibinfo {author} {\bibfnamefont {S.}~\bibnamefont
  {Goriely}},\ }\bibfield  {title} {\bibinfo {title} {Astronuclear physics: A
  tale of the atomic nuclei in the skies},\ }\href
  {https://doi.org/https://doi.org/10.1016/j.ppnp.2020.103766} {\bibfield
  {journal} {\bibinfo  {journal} {Progress in Particle and Nuclear Physics}\
  }\textbf {\bibinfo {volume} {112}},\ \bibinfo {pages} {103766} (\bibinfo
  {year} {2020})}\BibitemShut {NoStop}%
\bibitem [{\citenamefont {K\"appeler}\ \emph {et~al.}(2011)\citenamefont
  {K\"appeler}, \citenamefont {Gallino}, \citenamefont {Bisterzo},\ and\
  \citenamefont {Aoki}}]{Kaeppeler_2011}%
  \BibitemOpen
  \bibfield  {author} {\bibinfo {author} {\bibfnamefont {F.}~\bibnamefont
  {K\"appeler}}, \bibinfo {author} {\bibfnamefont {R.}~\bibnamefont {Gallino}},
  \bibinfo {author} {\bibfnamefont {S.}~\bibnamefont {Bisterzo}},\ and\
  \bibinfo {author} {\bibfnamefont {W.}~\bibnamefont {Aoki}},\ }\bibfield
  {title} {\bibinfo {title} {The $s$ process: Nuclear physics, stellar models,
  and observations},\ }\href {https://doi.org/10.1103/RevModPhys.83.157}
  {\bibfield  {journal} {\bibinfo  {journal} {Rev. Mod. Phys.}\ }\textbf
  {\bibinfo {volume} {83}},\ \bibinfo {pages} {157} (\bibinfo {year}
  {2011})}\BibitemShut {NoStop}%
\bibitem [{\citenamefont {Jaag}\ and\ \citenamefont
  {K\"appeler}(1996)}]{Jaag_1996}%
  \BibitemOpen
  \bibfield  {author} {\bibinfo {author} {\bibfnamefont {S.}~\bibnamefont
  {Jaag}}\ and\ \bibinfo {author} {\bibfnamefont {F.}~\bibnamefont
  {K\"appeler}},\ }\bibfield  {title} {\bibinfo {title} {{The Stellar
  $(n,\,\gamma)$ Cross Section of the Unstable Isotope $^{163}$Ho and the
  Origin of $^{164}$Er}},\ }\href {https://doi.org/10.1086/177375} {\bibfield
  {journal} {\bibinfo  {journal} {\apj}\ }\textbf {\bibinfo {volume} {464}},\
  \bibinfo {pages} {874} (\bibinfo {year} {1996})}\BibitemShut {NoStop}%
\bibitem [{\citenamefont {Hayakawa}\ \emph {et~al.}(2008)\citenamefont
  {Hayakawa}, \citenamefont {Shizuma}, \citenamefont {Miyamoto}, \citenamefont
  {Amano}, \citenamefont {Horikawa}, \citenamefont {Ishihara}, \citenamefont
  {Mori}, \citenamefont {Kawase}, \citenamefont {Kando}, \citenamefont
  {Kikuzawa}, \citenamefont {Chiba}, \citenamefont {Mochizuki}, \citenamefont
  {Kajino},\ and\ \citenamefont {Fujiwara}}]{Hayakawa_2008}%
  \BibitemOpen
  \bibfield  {author} {\bibinfo {author} {\bibfnamefont {T.}~\bibnamefont
  {Hayakawa}}, \bibinfo {author} {\bibfnamefont {T.}~\bibnamefont {Shizuma}},
  \bibinfo {author} {\bibfnamefont {S.}~\bibnamefont {Miyamoto}}, \bibinfo
  {author} {\bibfnamefont {S.}~\bibnamefont {Amano}}, \bibinfo {author}
  {\bibfnamefont {K.}~\bibnamefont {Horikawa}}, \bibinfo {author}
  {\bibfnamefont {K.}~\bibnamefont {Ishihara}}, \bibinfo {author}
  {\bibfnamefont {M.}~\bibnamefont {Mori}}, \bibinfo {author} {\bibfnamefont
  {K.}~\bibnamefont {Kawase}}, \bibinfo {author} {\bibfnamefont
  {M.}~\bibnamefont {Kando}}, \bibinfo {author} {\bibfnamefont
  {N.}~\bibnamefont {Kikuzawa}}, \bibinfo {author} {\bibfnamefont
  {S.}~\bibnamefont {Chiba}}, \bibinfo {author} {\bibfnamefont
  {T.}~\bibnamefont {Mochizuki}}, \bibinfo {author} {\bibfnamefont
  {T.}~\bibnamefont {Kajino}},\ and\ \bibinfo {author} {\bibfnamefont
  {M.}~\bibnamefont {Fujiwara}},\ }\bibfield  {title} {\bibinfo {title}
  {Half-life of the $^{164}\mathrm{Ho}$ by the ($\ensuremath{\gamma},n$)
  reaction from laser {Compton} scattering $\ensuremath{\gamma}$ rays at the
  electron storage ring {NewSUBARU}},\ }\href
  {https://doi.org/10.1103/PhysRevC.77.068801} {\bibfield  {journal} {\bibinfo
  {journal} {Phys. Rev. C}\ }\textbf {\bibinfo {volume} {77}},\ \bibinfo
  {pages} {068801} (\bibinfo {year} {2008})}\BibitemShut {NoStop}%
\bibitem [{\citenamefont {Pogliano}\ \emph {et~al.}(2023)\citenamefont
  {Pogliano}, \citenamefont {Larsen}, \citenamefont {Goriely}, \citenamefont
  {Siess}, \citenamefont {Markova}, \citenamefont {G\"orgen}, \citenamefont
  {Heines}, \citenamefont {Ingeberg}, \citenamefont {Kjus}, \citenamefont
  {Larsson}, \citenamefont {Li}, \citenamefont {Martinsen}, \citenamefont
  {Owens-Fryar}, \citenamefont {Pedersen}, \citenamefont {Siem}, \citenamefont
  {Torvund},\ and\ \citenamefont {Tsantiri}}]{Pogliano_2023}%
  \BibitemOpen
  \bibfield  {author} {\bibinfo {author} {\bibfnamefont {F.}~\bibnamefont
  {Pogliano}}, \bibinfo {author} {\bibfnamefont {A.~C.}\ \bibnamefont
  {Larsen}}, \bibinfo {author} {\bibfnamefont {S.}~\bibnamefont {Goriely}},
  \bibinfo {author} {\bibfnamefont {L.}~\bibnamefont {Siess}}, \bibinfo
  {author} {\bibfnamefont {M.}~\bibnamefont {Markova}}, \bibinfo {author}
  {\bibfnamefont {A.}~\bibnamefont {G\"orgen}}, \bibinfo {author}
  {\bibfnamefont {J.}~\bibnamefont {Heines}}, \bibinfo {author} {\bibfnamefont
  {V.~W.}\ \bibnamefont {Ingeberg}}, \bibinfo {author} {\bibfnamefont {R.~G.}\
  \bibnamefont {Kjus}}, \bibinfo {author} {\bibfnamefont {J.~E.~L.}\
  \bibnamefont {Larsson}}, \bibinfo {author} {\bibfnamefont {K.~C.~W.}\
  \bibnamefont {Li}}, \bibinfo {author} {\bibfnamefont {E.~M.}\ \bibnamefont
  {Martinsen}}, \bibinfo {author} {\bibfnamefont {G.~J.}\ \bibnamefont
  {Owens-Fryar}}, \bibinfo {author} {\bibfnamefont {L.~G.}\ \bibnamefont
  {Pedersen}}, \bibinfo {author} {\bibfnamefont {S.}~\bibnamefont {Siem}},
  \bibinfo {author} {\bibfnamefont {G.~S.}\ \bibnamefont {Torvund}},\ and\
  \bibinfo {author} {\bibfnamefont {A.}~\bibnamefont {Tsantiri}},\ }\bibfield
  {title} {\bibinfo {title} {Experimentally constrained
  $^{165,166}\mathrm{Ho}(n,\ensuremath{\gamma})$ rates and implications for the
  $s$ process},\ }\href {https://doi.org/10.1103/PhysRevC.107.064614}
  {\bibfield  {journal} {\bibinfo  {journal} {Phys. Rev. C}\ }\textbf {\bibinfo
  {volume} {107}},\ \bibinfo {pages} {064614} (\bibinfo {year}
  {2023})}\BibitemShut {NoStop}%
\bibitem [{\citenamefont {Martinet}\ \emph {et~al.}(2024)\citenamefont
  {Martinet}, \citenamefont {Choplin}, \citenamefont {Goriely},\ and\
  \citenamefont {Siess}}]{Martinet_2024}%
  \BibitemOpen
  \bibfield  {author} {\bibinfo {author} {\bibfnamefont {S.}~\bibnamefont
  {Martinet}}, \bibinfo {author} {\bibfnamefont {A.}~\bibnamefont {Choplin}},
  \bibinfo {author} {\bibfnamefont {S.}~\bibnamefont {Goriely}},\ and\ \bibinfo
  {author} {\bibfnamefont {L.}~\bibnamefont {Siess}},\ }\bibfield  {title}
  {\bibinfo {title} {The intermediate neutron capture process. {IV. Impact} of
  nuclear model and parameter uncertainties},\ }\href
  {https://doi.org/10.1051/0004-6361/202347734} {\bibfield  {journal} {\bibinfo
   {journal} {Astronomy and Astrophysics}\ }\textbf {\bibinfo {volume} {684}},\
  \bibinfo {pages} {A8} (\bibinfo {year} {2024})}\BibitemShut {NoStop}%
\bibitem [{\citenamefont {Reifarth}\ \emph {et~al.}(2003)\citenamefont
  {Reifarth}, \citenamefont {Haight}, \citenamefont {Heil}, \citenamefont
  {Fowler}, \citenamefont {K\"appeler}, \citenamefont {Miller}, \citenamefont
  {Rundberg}, \citenamefont {Ullmann},\ and\ \citenamefont
  {Wilhelmy}}]{Reifarth_2003}%
  \BibitemOpen
  \bibfield  {author} {\bibinfo {author} {\bibfnamefont {R.}~\bibnamefont
  {Reifarth}}, \bibinfo {author} {\bibfnamefont {R.}~\bibnamefont {Haight}},
  \bibinfo {author} {\bibfnamefont {M.}~\bibnamefont {Heil}}, \bibinfo {author}
  {\bibfnamefont {M.}~\bibnamefont {Fowler}}, \bibinfo {author} {\bibfnamefont
  {F.}~\bibnamefont {K\"appeler}}, \bibinfo {author} {\bibfnamefont
  {G.}~\bibnamefont {Miller}}, \bibinfo {author} {\bibfnamefont
  {R.}~\bibnamefont {Rundberg}}, \bibinfo {author} {\bibfnamefont
  {J.}~\bibnamefont {Ullmann}},\ and\ \bibinfo {author} {\bibfnamefont
  {J.}~\bibnamefont {Wilhelmy}},\ }\bibfield  {title} {\bibinfo {title}
  {Neutron capture measurements on $^{171}${Tm}},\ }\href
  {https://doi.org/https://doi.org/10.1016/S0375-9474(03)00862-5} {\bibfield
  {journal} {\bibinfo  {journal} {Nuclear Physics A}\ }\textbf {\bibinfo
  {volume} {718}},\ \bibinfo {pages} {478} (\bibinfo {year}
  {2003})}\BibitemShut {NoStop}%
\bibitem [{\citenamefont {Guerrero}\ \emph {et~al.}(2020)\citenamefont
  {Guerrero}, \citenamefont {Lerendegui-Marco}, \citenamefont {Paul},
  \citenamefont {Tessler}, \citenamefont {Heinitz}, \citenamefont
  {Domingo-Pardo}, \citenamefont {Cristallo}, \citenamefont {Dressler},
  \citenamefont {Halfon}, \citenamefont {Kivel}, \citenamefont {K\"oster},
  \citenamefont {Maugeri}, \citenamefont {Palchan-Hazan}, \citenamefont
  {Quesada}, \citenamefont {Rochman},\ and\ \citenamefont {\emph{et
  al.}}}]{Guerrero_2020}%
  \BibitemOpen
  \bibfield  {author} {\bibinfo {author} {\bibfnamefont {C.}~\bibnamefont
  {Guerrero}}, \bibinfo {author} {\bibfnamefont {J.}~\bibnamefont
  {Lerendegui-Marco}}, \bibinfo {author} {\bibfnamefont {M.}~\bibnamefont
  {Paul}}, \bibinfo {author} {\bibfnamefont {M.}~\bibnamefont {Tessler}},
  \bibinfo {author} {\bibfnamefont {S.}~\bibnamefont {Heinitz}}, \bibinfo
  {author} {\bibfnamefont {C.}~\bibnamefont {Domingo-Pardo}}, \bibinfo {author}
  {\bibfnamefont {S.}~\bibnamefont {Cristallo}}, \bibinfo {author}
  {\bibfnamefont {R.}~\bibnamefont {Dressler}}, \bibinfo {author}
  {\bibfnamefont {S.}~\bibnamefont {Halfon}}, \bibinfo {author} {\bibfnamefont
  {N.}~\bibnamefont {Kivel}}, \bibinfo {author} {\bibfnamefont
  {U.}~\bibnamefont {K\"oster}}, \bibinfo {author} {\bibfnamefont {E.~A.}\
  \bibnamefont {Maugeri}}, \bibinfo {author} {\bibfnamefont {T.}~\bibnamefont
  {Palchan-Hazan}}, \bibinfo {author} {\bibfnamefont {J.~M.}\ \bibnamefont
  {Quesada}}, \bibinfo {author} {\bibfnamefont {D.}~\bibnamefont {Rochman}},\
  and\ \bibinfo {author} {\bibnamefont {\emph{et al.}}} (\bibinfo
  {collaboration} {nTOF Collaboration}),\ }\bibfield  {title} {\bibinfo {title}
  {Neutron capture on the $s$-process branching point $^{171}\mathrm{Tm}$ via
  time-of-flight and activation},\ }\href
  {https://doi.org/10.1103/PhysRevLett.125.142701} {\bibfield  {journal}
  {\bibinfo  {journal} {Phys. Rev. Lett.}\ }\textbf {\bibinfo {volume} {125}},\
  \bibinfo {pages} {142701} (\bibinfo {year} {2020})}\BibitemShut {NoStop}%
\bibitem [{\citenamefont {Wolynec}\ and\ \citenamefont
  {Martins}(1987)}]{wolynec_1987}%
  \BibitemOpen
  \bibfield  {author} {\bibinfo {author} {\bibfnamefont {E.}~\bibnamefont
  {Wolynec}}\ and\ \bibinfo {author} {\bibfnamefont {M.~N.}\ \bibnamefont
  {Martins}},\ }\bibfield  {title} {\bibinfo {title} {Discrepancies between
  {Saclay} and {Livermore} photoneutron cross sections},\ }\href
  {https://sbfisica.org.br/bjp/download/v17/v17a05.pdf} {\bibfield  {journal}
  {\bibinfo  {journal} {Revista Brasileira de Fisica}\ }\textbf {\bibinfo
  {volume} {17}},\ \bibinfo {pages} {56} (\bibinfo {year} {1987})}\BibitemShut
  {NoStop}%
\bibitem [{\citenamefont {Varlamov}\ \emph {et~al.}(2014)\citenamefont
  {Varlamov}, \citenamefont {Ishkhanov}, \citenamefont {Orlin},\ and\
  \citenamefont {Stopani}}]{varlamov_2014}%
  \BibitemOpen
  \bibfield  {author} {\bibinfo {author} {\bibfnamefont {V.~V.}\ \bibnamefont
  {Varlamov}}, \bibinfo {author} {\bibfnamefont {B.~S.}\ \bibnamefont
  {Ishkhanov}}, \bibinfo {author} {\bibfnamefont {V.~N.}\ \bibnamefont
  {Orlin}},\ and\ \bibinfo {author} {\bibfnamefont {K.~A.}\ \bibnamefont
  {Stopani}},\ }\bibfield  {title} {\bibinfo {title} {A new approach for
  analysis and evaluation of partial photoneutron reaction cross sections},\
  }\href {https://doi.org/10.1140/epja/i2014-14114-x} {\bibfield  {journal}
  {\bibinfo  {journal} {The European Physical Journal A}\ }\textbf {\bibinfo
  {volume} {50}},\ \bibinfo {pages} {114} (\bibinfo {year} {2014})}\BibitemShut
  {NoStop}%
\bibitem [{\citenamefont {Wolsey}(2019)}]{Wolsey_2019}%
  \BibitemOpen
  \bibfield  {author} {\bibinfo {author} {\bibfnamefont {K.}~\bibnamefont
  {Wolsey}},\ }\emph {\bibinfo {title} {Measurements of the
  $^{124}$Sn$(\gamma,\,n)$ and $^{169}$Tm$(\gamma,\,n)$ cross sections at an
  incident photon energy of 13~MeV}},\ \href@noop {} {\bibinfo {type} {B.{Sc}.
  thesis}},\ \bibinfo  {school} {Department of Physics, Brigham Young
  University - Idaho} (\bibinfo {year} {2019})\BibitemShut {NoStop}%
\bibitem [{\citenamefont {Filipescu}\ \emph {et~al.}(2023)\citenamefont
  {Filipescu}, \citenamefont {Gheorghe}, \citenamefont {Stopani}, \citenamefont
  {Belyshev}, \citenamefont {Hashimoto}, \citenamefont {Miyamoto},\ and\
  \citenamefont {Utsunomiya}}]{filipescu_2023}%
  \BibitemOpen
  \bibfield  {author} {\bibinfo {author} {\bibfnamefont {D.}~\bibnamefont
  {Filipescu}}, \bibinfo {author} {\bibfnamefont {I.}~\bibnamefont {Gheorghe}},
  \bibinfo {author} {\bibfnamefont {K.}~\bibnamefont {Stopani}}, \bibinfo
  {author} {\bibfnamefont {S.}~\bibnamefont {Belyshev}}, \bibinfo {author}
  {\bibfnamefont {S.}~\bibnamefont {Hashimoto}}, \bibinfo {author}
  {\bibfnamefont {S.}~\bibnamefont {Miyamoto}},\ and\ \bibinfo {author}
  {\bibfnamefont {H.}~\bibnamefont {Utsunomiya}},\ }\bibfield  {title}
  {\bibinfo {title} {Spectral distribution and flux of $\gamma$-ray beams
  produced through {Compton} scattering of unsynchronized laser and electron
  beams},\ }\href {https://doi.org/https://doi.org/10.1016/j.nima.2022.167885}
  {\bibfield  {journal} {\bibinfo  {journal} {Nuclear Instruments and Methods
  in Physics Research Section A: Accelerators, Spectrometers, Detectors and
  Associated Equipment}\ }\textbf {\bibinfo {volume} {1047}},\ \bibinfo {pages}
  {167885} (\bibinfo {year} {2023})},\ \bibinfo {note}
  {arXiv:2211.14650}\BibitemShut {NoStop}%
\bibitem [{\citenamefont {Utsunomiya}\ \emph {et~al.}(2018)\citenamefont
  {Utsunomiya}, \citenamefont {Watanabe}, \citenamefont {Ari-izumi},
  \citenamefont {Takenaka}, \citenamefont {Araki}, \citenamefont {Tsuji},
  \citenamefont {Gheorghe}, \citenamefont {Filipescu}, \citenamefont
  {Belyshev}, \citenamefont {Stopani}, \citenamefont {Symochko}, \citenamefont
  {Wang}, \citenamefont {Fan}, \citenamefont {Renstr\o{}m}, \citenamefont
  {Tveten}, \citenamefont {Lui}, \citenamefont {Sugita},\ and\ \citenamefont
  {Miyamoto}}]{utsunomiya_2018}%
  \BibitemOpen
  \bibfield  {author} {\bibinfo {author} {\bibfnamefont {H.}~\bibnamefont
  {Utsunomiya}}, \bibinfo {author} {\bibfnamefont {T.}~\bibnamefont
  {Watanabe}}, \bibinfo {author} {\bibfnamefont {T.}~\bibnamefont {Ari-izumi}},
  \bibinfo {author} {\bibfnamefont {D.}~\bibnamefont {Takenaka}}, \bibinfo
  {author} {\bibfnamefont {T.}~\bibnamefont {Araki}}, \bibinfo {author}
  {\bibfnamefont {K.}~\bibnamefont {Tsuji}}, \bibinfo {author} {\bibfnamefont
  {I.}~\bibnamefont {Gheorghe}}, \bibinfo {author} {\bibfnamefont {D.~M.}\
  \bibnamefont {Filipescu}}, \bibinfo {author} {\bibfnamefont {S.}~\bibnamefont
  {Belyshev}}, \bibinfo {author} {\bibfnamefont {K.}~\bibnamefont {Stopani}},
  \bibinfo {author} {\bibfnamefont {D.}~\bibnamefont {Symochko}}, \bibinfo
  {author} {\bibfnamefont {H.}~\bibnamefont {Wang}}, \bibinfo {author}
  {\bibfnamefont {G.}~\bibnamefont {Fan}}, \bibinfo {author} {\bibfnamefont
  {T.}~\bibnamefont {Renstr\o{}m}}, \bibinfo {author} {\bibfnamefont {G.~M.}\
  \bibnamefont {Tveten}}, \bibinfo {author} {\bibfnamefont {Y.-W.}\
  \bibnamefont {Lui}}, \bibinfo {author} {\bibfnamefont {K.}~\bibnamefont
  {Sugita}},\ and\ \bibinfo {author} {\bibfnamefont {S.}~\bibnamefont
  {Miyamoto}},\ }\bibfield  {title} {\bibinfo {title} {Photon-flux
  determination by the {Poisson-fitting} technique with quenching
  corrections},\ }\href
  {https://doi.org/https://doi.org/10.1016/j.nima.2018.04.021} {\bibfield
  {journal} {\bibinfo  {journal} {Nuclear Instruments and Methods in Physics
  Research Section A: Accelerators, Spectrometers, Detectors and Associated
  Equipment}\ }\textbf {\bibinfo {volume} {896}},\ \bibinfo {pages} {103}
  (\bibinfo {year} {2018})}\BibitemShut {NoStop}%
\bibitem [{\citenamefont {Gheorghe}\ \emph {et~al.}(2021)\citenamefont
  {Gheorghe}, \citenamefont {Utsunomiya}, \citenamefont {Stopani},
  \citenamefont {Filipescu}, \citenamefont {Ari-izumi}, \citenamefont
  {Belyshev}, \citenamefont {Fan}, \citenamefont {Krzysiek}, \citenamefont
  {Liu}, \citenamefont {Lui}, \citenamefont {Symochko}, \citenamefont {Wang},\
  and\ \citenamefont {Miyamoto}}]{gheorghe_2021}%
  \BibitemOpen
  \bibfield  {author} {\bibinfo {author} {\bibfnamefont {I.}~\bibnamefont
  {Gheorghe}}, \bibinfo {author} {\bibfnamefont {H.}~\bibnamefont
  {Utsunomiya}}, \bibinfo {author} {\bibfnamefont {K.}~\bibnamefont {Stopani}},
  \bibinfo {author} {\bibfnamefont {D.}~\bibnamefont {Filipescu}}, \bibinfo
  {author} {\bibfnamefont {T.}~\bibnamefont {Ari-izumi}}, \bibinfo {author}
  {\bibfnamefont {S.}~\bibnamefont {Belyshev}}, \bibinfo {author}
  {\bibfnamefont {G.}~\bibnamefont {Fan}}, \bibinfo {author} {\bibfnamefont
  {M.}~\bibnamefont {Krzysiek}}, \bibinfo {author} {\bibfnamefont
  {L.}~\bibnamefont {Liu}}, \bibinfo {author} {\bibfnamefont {Y.-W.}\
  \bibnamefont {Lui}}, \bibinfo {author} {\bibfnamefont {D.}~\bibnamefont
  {Symochko}}, \bibinfo {author} {\bibfnamefont {H.}~\bibnamefont {Wang}},\
  and\ \bibinfo {author} {\bibfnamefont {S.}~\bibnamefont {Miyamoto}},\
  }\bibfield  {title} {\bibinfo {title} {Updated neutron-multiplicity sorting
  method for producing photoneutron average energies and resolving multiple
  firing events},\ }\href
  {https://doi.org/https://doi.org/10.1016/j.nima.2021.165867} {\bibfield
  {journal} {\bibinfo  {journal} {Nuclear Instruments and Methods in Physics
  Research Section A: Accelerators, Spectrometers, Detectors and Associated
  Equipment}\ }\textbf {\bibinfo {volume} {1019}},\ \bibinfo {pages} {165867}
  (\bibinfo {year} {2021})}\BibitemShut {NoStop}%
\bibitem [{\citenamefont {Hashimoto}\ and\ \citenamefont
  {Hirakawa}(2023)}]{hashimoto_2022}%
  \BibitemOpen
  \bibfield  {author} {\bibinfo {author} {\bibfnamefont {S.}~\bibnamefont
  {Hashimoto}}\ and\ \bibinfo {author} {\bibfnamefont {H.}~\bibnamefont
  {Hirakawa}},\ }\bibinfo {title} {Increased $\mathrm{LCS}$ $\gamma$-ray flux
  by optimizing laser optics at $\mathrm{BL01}$},\ in\ \href@noop {} {\emph
  {\bibinfo {booktitle} {Laboratory of Advanced Science and Technology for
  Industry, University of Hyogo, LASTI Annual Report}}},\ Vol.~\bibinfo
  {volume} {24}\ (\bibinfo {year} {2023})\ p.~\bibinfo {pages} {17}\BibitemShut
  {NoStop}%
\bibitem [{\citenamefont {Utsunomiya}\ \emph {et~al.}(2014)\citenamefont
  {Utsunomiya}, \citenamefont {Shima}, \citenamefont {Takahisa}, \citenamefont
  {Filipescu}, \citenamefont {Tesileanu}, \citenamefont {Gheorghe},
  \citenamefont {Nyhus}, \citenamefont {Renstr\o{}m}, \citenamefont {Lui},
  \citenamefont {Kitagawa}, \citenamefont {Amano},\ and\ \citenamefont
  {Miyamoto}}]{utsunomiya_2014}%
  \BibitemOpen
  \bibfield  {author} {\bibinfo {author} {\bibfnamefont {H.}~\bibnamefont
  {Utsunomiya}}, \bibinfo {author} {\bibfnamefont {T.}~\bibnamefont {Shima}},
  \bibinfo {author} {\bibfnamefont {K.}~\bibnamefont {Takahisa}}, \bibinfo
  {author} {\bibfnamefont {D.~M.}\ \bibnamefont {Filipescu}}, \bibinfo {author}
  {\bibfnamefont {O.}~\bibnamefont {Tesileanu}}, \bibinfo {author}
  {\bibfnamefont {I.}~\bibnamefont {Gheorghe}}, \bibinfo {author}
  {\bibfnamefont {H.-T.}\ \bibnamefont {Nyhus}}, \bibinfo {author}
  {\bibfnamefont {T.}~\bibnamefont {Renstr\o{}m}}, \bibinfo {author}
  {\bibfnamefont {Y.-W.}\ \bibnamefont {Lui}}, \bibinfo {author} {\bibfnamefont
  {Y.}~\bibnamefont {Kitagawa}}, \bibinfo {author} {\bibfnamefont
  {S.}~\bibnamefont {Amano}},\ and\ \bibinfo {author} {\bibfnamefont
  {S.}~\bibnamefont {Miyamoto}},\ }\bibfield  {title} {\bibinfo {title}
  {{Energy Calibration of the NewSUBARU Storage Ring for Laser
  Compton-Scattering Gamma Rays and Applications}},\ }\href
  {https://doi.org/10.1109/TNS.2014.2312323} {\bibfield  {journal} {\bibinfo
  {journal} {IEEE Transactions on Nuclear Science}\ }\textbf {\bibinfo {volume}
  {61}},\ \bibinfo {pages} {1252} (\bibinfo {year} {2014})}\BibitemShut
  {NoStop}%
\bibitem [{\citenamefont {Ari-Izumi}\ \emph {et~al.}(2023)\citenamefont
  {Ari-Izumi}, \citenamefont {Gheorghe}, \citenamefont {Filipescu},
  \citenamefont {Hashimoto}, \citenamefont {Miyamoto},\ and\ \citenamefont
  {Utsunomiya}}]{Ariizumi_2023}%
  \BibitemOpen
  \bibfield  {author} {\bibinfo {author} {\bibfnamefont {T.}~\bibnamefont
  {Ari-Izumi}}, \bibinfo {author} {\bibfnamefont {I.}~\bibnamefont {Gheorghe}},
  \bibinfo {author} {\bibfnamefont {D.}~\bibnamefont {Filipescu}}, \bibinfo
  {author} {\bibfnamefont {S.}~\bibnamefont {Hashimoto}}, \bibinfo {author}
  {\bibfnamefont {S.}~\bibnamefont {Miyamoto}},\ and\ \bibinfo {author}
  {\bibfnamefont {H.}~\bibnamefont {Utsunomiya}},\ }\bibfield  {title}
  {\bibinfo {title} {{Spatial profiles of collimated laser Compton-scattering
  $\gamma$-ray beams}},\ }\href
  {https://doi.org/10.1088/1748-0221/18/06/T06005} {\bibfield  {journal}
  {\bibinfo  {journal} {Journal of Instrumentation}\ }\textbf {\bibinfo
  {volume} {18}}\bibinfo  {number} { (06)},\ \bibinfo {pages}
  {T06005}}\BibitemShut {NoStop}%
\bibitem [{\citenamefont {Filipescu}(2022)}]{filipescu_2022}%
  \BibitemOpen
\bibfield  {number} {  }\bibfield  {author} {\bibinfo {author} {\bibfnamefont
  {D.}~\bibnamefont {Filipescu}},\ }\bibfield  {title} {\bibinfo {title}
  {{Monte Carlo simulation method of polarization effects in Laser Compton
  Scattering on relativistic electrons}},\ }\href
  {https://doi.org/10.1088/1748-0221/17/11/P11006} {\bibfield  {journal}
  {\bibinfo  {journal} {Journal of Instrumentation}\ }\textbf {\bibinfo
  {volume} {17}}\bibinfo  {number} { (11)},\ \bibinfo {pages}
  {P11006}}\BibitemShut {NoStop}%
\bibitem [{\citenamefont {Utsunomiya}\ \emph {et~al.}(2015)\citenamefont
  {Utsunomiya}, \citenamefont {Katayama}, \citenamefont {Gheorghe},
  \citenamefont {Imai}, \citenamefont {Yamaguchi}, \citenamefont {Kahl},
  \citenamefont {Sakaguchi}, \citenamefont {Shima}, \citenamefont {Takahisa},\
  and\ \citenamefont {Miyamoto}}]{utsunomiya_2015}%
  \BibitemOpen
\bibfield  {number} {  }\bibfield  {author} {\bibinfo {author} {\bibfnamefont
  {H.}~\bibnamefont {Utsunomiya}}, \bibinfo {author} {\bibfnamefont
  {S.}~\bibnamefont {Katayama}}, \bibinfo {author} {\bibfnamefont
  {I.}~\bibnamefont {Gheorghe}}, \bibinfo {author} {\bibfnamefont
  {S.}~\bibnamefont {Imai}}, \bibinfo {author} {\bibfnamefont {H.}~\bibnamefont
  {Yamaguchi}}, \bibinfo {author} {\bibfnamefont {D.}~\bibnamefont {Kahl}},
  \bibinfo {author} {\bibfnamefont {Y.}~\bibnamefont {Sakaguchi}}, \bibinfo
  {author} {\bibfnamefont {T.}~\bibnamefont {Shima}}, \bibinfo {author}
  {\bibfnamefont {K.}~\bibnamefont {Takahisa}},\ and\ \bibinfo {author}
  {\bibfnamefont {S.}~\bibnamefont {Miyamoto}},\ }\bibfield  {title} {\bibinfo
  {title} {Photodisintegration of $^{9}\mathrm{Be}$ through the $1/{2}^{+}$
  state and cluster dipole resonance},\ }\href
  {https://doi.org/10.1103/PhysRevC.92.064323} {\bibfield  {journal} {\bibinfo
  {journal} {Phys. Rev. C}\ }\textbf {\bibinfo {volume} {92}},\ \bibinfo
  {pages} {064323} (\bibinfo {year} {2015})}\BibitemShut {NoStop}%
\bibitem [{\citenamefont {Utsunomiya}\ \emph {et~al.}(2017)\citenamefont
  {Utsunomiya}, \citenamefont {Gheorghe}, \citenamefont {Filipescu},
  \citenamefont {Glodariu}, \citenamefont {Belyshev}, \citenamefont {Stopani},
  \citenamefont {Varlamov}, \citenamefont {Ishkhanov}, \citenamefont
  {Katayama}, \citenamefont {Takenaka}, \citenamefont {Ari-izumi},
  \citenamefont {Amano},\ and\ \citenamefont {Miyamoto}}]{utsunomiya_2017}%
  \BibitemOpen
  \bibfield  {author} {\bibinfo {author} {\bibfnamefont {H.}~\bibnamefont
  {Utsunomiya}}, \bibinfo {author} {\bibfnamefont {I.}~\bibnamefont
  {Gheorghe}}, \bibinfo {author} {\bibfnamefont {D.~M.}\ \bibnamefont
  {Filipescu}}, \bibinfo {author} {\bibfnamefont {T.}~\bibnamefont {Glodariu}},
  \bibinfo {author} {\bibfnamefont {S.}~\bibnamefont {Belyshev}}, \bibinfo
  {author} {\bibfnamefont {K.}~\bibnamefont {Stopani}}, \bibinfo {author}
  {\bibfnamefont {V.}~\bibnamefont {Varlamov}}, \bibinfo {author}
  {\bibfnamefont {B.}~\bibnamefont {Ishkhanov}}, \bibinfo {author}
  {\bibfnamefont {S.}~\bibnamefont {Katayama}}, \bibinfo {author}
  {\bibfnamefont {D.}~\bibnamefont {Takenaka}}, \bibinfo {author}
  {\bibfnamefont {T.}~\bibnamefont {Ari-izumi}}, \bibinfo {author}
  {\bibfnamefont {S.}~\bibnamefont {Amano}},\ and\ \bibinfo {author}
  {\bibfnamefont {S.}~\bibnamefont {Miyamoto}},\ }\bibfield  {title} {\bibinfo
  {title} {Direct neutron-multiplicity sorting with a flat-efficiency
  detector},\ }\href
  {https://doi.org/https://doi.org/10.1016/j.nima.2017.08.001} {\bibfield
  {journal} {\bibinfo  {journal} {Nuclear Instruments and Methods in Physics
  Research Section A: Accelerators, Spectrometers, Detectors and Associated
  Equipment}\ }\textbf {\bibinfo {volume} {871}},\ \bibinfo {pages} {135}
  (\bibinfo {year} {2017})}\BibitemShut {NoStop}%
\bibitem [{\citenamefont {Renstr\o{}m}\ \emph {et~al.}(2018)\citenamefont
  {Renstr\o{}m}, \citenamefont {Utsunomiya}, \citenamefont {Nyhus},
  \citenamefont {Larsen}, \citenamefont {Guttormsen}, \citenamefont {Tveten},
  \citenamefont {Filipescu}, \citenamefont {Gheorghe}, \citenamefont {Goriely},
  \citenamefont {Hilaire}, \citenamefont {Lui}, \citenamefont {Midtb\o{}},
  \citenamefont {P\'eru}, \citenamefont {Shima}, \citenamefont {Siem},\ and\
  \citenamefont {Tesileanu}}]{Renstrom18}%
  \BibitemOpen
  \bibfield  {author} {\bibinfo {author} {\bibfnamefont {T.}~\bibnamefont
  {Renstr\o{}m}}, \bibinfo {author} {\bibfnamefont {H.}~\bibnamefont
  {Utsunomiya}}, \bibinfo {author} {\bibfnamefont {H.~T.}\ \bibnamefont
  {Nyhus}}, \bibinfo {author} {\bibfnamefont {A.~C.}\ \bibnamefont {Larsen}},
  \bibinfo {author} {\bibfnamefont {M.}~\bibnamefont {Guttormsen}}, \bibinfo
  {author} {\bibfnamefont {G.~M.}\ \bibnamefont {Tveten}}, \bibinfo {author}
  {\bibfnamefont {D.~M.}\ \bibnamefont {Filipescu}}, \bibinfo {author}
  {\bibfnamefont {I.}~\bibnamefont {Gheorghe}}, \bibinfo {author}
  {\bibfnamefont {S.}~\bibnamefont {Goriely}}, \bibinfo {author} {\bibfnamefont
  {S.}~\bibnamefont {Hilaire}}, \bibinfo {author} {\bibfnamefont {Y.-W.}\
  \bibnamefont {Lui}}, \bibinfo {author} {\bibfnamefont {J.~E.}\ \bibnamefont
  {Midtb\o{}}}, \bibinfo {author} {\bibfnamefont {S.}~\bibnamefont {P\'eru}},
  \bibinfo {author} {\bibfnamefont {T.}~\bibnamefont {Shima}}, \bibinfo
  {author} {\bibfnamefont {S.}~\bibnamefont {Siem}},\ and\ \bibinfo {author}
  {\bibfnamefont {O.}~\bibnamefont {Tesileanu}},\ }\bibfield  {title} {\bibinfo
  {title} {{Verification of detailed balance for $\ensuremath{\gamma}$
  absorption and emission in Dy isotopes}},\ }\href
  {https://doi.org/10.1103/PhysRevC.98.054310} {\bibfield  {journal} {\bibinfo
  {journal} {Phys. Rev. C}\ }\textbf {\bibinfo {volume} {98}},\ \bibinfo
  {pages} {054310} (\bibinfo {year} {2018})}\BibitemShut {NoStop}%
\bibitem [{\citenamefont {Larsen}\ \emph {et~al.}(2023)\citenamefont {Larsen},
  \citenamefont {Tveten}, \citenamefont {Renstr\o{}m}, \citenamefont
  {Utsunomiya}, \citenamefont {Algin}, \citenamefont {Ari-izumi}, \citenamefont
  {Ay}, \citenamefont {Bello~Garrote}, \citenamefont {Crespo~Campo},
  \citenamefont {Furmyr}, \citenamefont {Goriely}, \citenamefont {G\"orgen},
  \citenamefont {Guttormsen}, \citenamefont {Ingeberg}, \citenamefont {Kheswa},
  \citenamefont {Kullmann}, \citenamefont {Laplace}, \citenamefont {Lima},
  \citenamefont {Markova}, \citenamefont {Midtb\o{}}, \citenamefont {Miyamoto},
  \citenamefont {Mj\o{}s}, \citenamefont {Modamio}, \citenamefont {Ozgur},
  \citenamefont {Pogliano}, \citenamefont {Riemer-S\o{}rensen}, \citenamefont
  {Sahin}, \citenamefont {Shen}, \citenamefont {Siem}, \citenamefont {Spyrou},\
  and\ \citenamefont {Wiedeking}}]{LarsenTveten23}%
  \BibitemOpen
  \bibfield  {author} {\bibinfo {author} {\bibfnamefont {A.~C.}\ \bibnamefont
  {Larsen}}, \bibinfo {author} {\bibfnamefont {G.~M.}\ \bibnamefont {Tveten}},
  \bibinfo {author} {\bibfnamefont {T.}~\bibnamefont {Renstr\o{}m}}, \bibinfo
  {author} {\bibfnamefont {H.}~\bibnamefont {Utsunomiya}}, \bibinfo {author}
  {\bibfnamefont {E.}~\bibnamefont {Algin}}, \bibinfo {author} {\bibfnamefont
  {T.}~\bibnamefont {Ari-izumi}}, \bibinfo {author} {\bibfnamefont {K.~O.}\
  \bibnamefont {Ay}}, \bibinfo {author} {\bibfnamefont {F.~L.}\ \bibnamefont
  {Bello~Garrote}}, \bibinfo {author} {\bibfnamefont {L.}~\bibnamefont
  {Crespo~Campo}}, \bibinfo {author} {\bibfnamefont {F.}~\bibnamefont
  {Furmyr}}, \bibinfo {author} {\bibfnamefont {S.}~\bibnamefont {Goriely}},
  \bibinfo {author} {\bibfnamefont {A.}~\bibnamefont {G\"orgen}}, \bibinfo
  {author} {\bibfnamefont {M.}~\bibnamefont {Guttormsen}}, \bibinfo {author}
  {\bibfnamefont {V.~W.}\ \bibnamefont {Ingeberg}}, \bibinfo {author}
  {\bibfnamefont {B.~V.}\ \bibnamefont {Kheswa}}, \bibinfo {author}
  {\bibfnamefont {I.~K.~B.}\ \bibnamefont {Kullmann}}, \bibinfo {author}
  {\bibfnamefont {T.}~\bibnamefont {Laplace}}, \bibinfo {author} {\bibfnamefont
  {E.}~\bibnamefont {Lima}}, \bibinfo {author} {\bibfnamefont {M.}~\bibnamefont
  {Markova}}, \bibinfo {author} {\bibfnamefont {J.~E.}\ \bibnamefont
  {Midtb\o{}}}, \bibinfo {author} {\bibfnamefont {S.}~\bibnamefont {Miyamoto}},
  \bibinfo {author} {\bibfnamefont {A.~H.}\ \bibnamefont {Mj\o{}s}}, \bibinfo
  {author} {\bibfnamefont {V.}~\bibnamefont {Modamio}}, \bibinfo {author}
  {\bibfnamefont {M.}~\bibnamefont {Ozgur}}, \bibinfo {author} {\bibfnamefont
  {F.}~\bibnamefont {Pogliano}}, \bibinfo {author} {\bibfnamefont
  {S.}~\bibnamefont {Riemer-S\o{}rensen}}, \bibinfo {author} {\bibfnamefont
  {E.}~\bibnamefont {Sahin}}, \bibinfo {author} {\bibfnamefont
  {S.}~\bibnamefont {Shen}}, \bibinfo {author} {\bibfnamefont {S.}~\bibnamefont
  {Siem}}, \bibinfo {author} {\bibfnamefont {A.}~\bibnamefont {Spyrou}},\ and\
  \bibinfo {author} {\bibfnamefont {M.}~\bibnamefont {Wiedeking}},\ }\bibfield
  {title} {\bibinfo {title} {New experimental constraint on the
  $^{185}\mathrm{W}(n,\ensuremath{\gamma})^{186}\mathrm{W}$ cross section},\
  }\href {https://doi.org/10.1103/PhysRevC.108.025804} {\bibfield  {journal}
  {\bibinfo  {journal} {Phys. Rev. C}\ }\textbf {\bibinfo {volume} {108}},\
  \bibinfo {pages} {025804} (\bibinfo {year} {2023})},\ \bibinfo {note}
  {arXiv:2301.13301}\BibitemShut {NoStop}%
\bibitem [{\citenamefont {Berg\`ere}\ \emph {et~al.}(1968)\citenamefont
  {Berg\`ere}, \citenamefont {Beil},\ and\ \citenamefont
  {Veyssi\`ere}}]{Bergere_1968}%
  \BibitemOpen
  \bibfield  {author} {\bibinfo {author} {\bibfnamefont {R.}~\bibnamefont
  {Berg\`ere}}, \bibinfo {author} {\bibfnamefont {H.}~\bibnamefont {Beil}},\
  and\ \bibinfo {author} {\bibfnamefont {A.}~\bibnamefont {Veyssi\`ere}},\
  }\bibfield  {title} {\bibinfo {title} {{Photoneutron cross sections of La,
  Tb, Ho and Ta}},\ }\href
  {https://doi.org/https://doi.org/10.1016/0375-9474(68)90433-8} {\bibfield
  {journal} {\bibinfo  {journal} {Nuclear Physics A}\ }\textbf {\bibinfo
  {volume} {121}},\ \bibinfo {pages} {463} (\bibinfo {year}
  {1968})}\BibitemShut {NoStop}%
\bibitem [{\citenamefont {Berman}\ \emph {et~al.}(1969)\citenamefont {Berman},
  \citenamefont {Kelly}, \citenamefont {Bramblett†}, \citenamefont
  {Caldwell}, \citenamefont {Davis},\ and\ \citenamefont
  {Fultz}}]{Berman_1969}%
  \BibitemOpen
  \bibfield  {author} {\bibinfo {author} {\bibfnamefont {B.}~\bibnamefont
  {Berman}}, \bibinfo {author} {\bibfnamefont {M.}~\bibnamefont {Kelly}},
  \bibinfo {author} {\bibfnamefont {R.}~\bibnamefont {Bramblett†}}, \bibinfo
  {author} {\bibfnamefont {J.}~\bibnamefont {Caldwell}}, \bibinfo {author}
  {\bibfnamefont {H.}~\bibnamefont {Davis}},\ and\ \bibinfo {author}
  {\bibfnamefont {S.}~\bibnamefont {Fultz}},\ }\bibfield  {title} {\bibinfo
  {title} {{Giant Resonance in Deformed Nuclei: Photoneutron Cross Sections for
  Eu$^{153}$, Gd$^{160}$, Ho$^{165}$, and W$^{186}$}},\ }\href
  {https://doi.org/10.1103/PhysRev.185.1576} {\bibfield  {journal} {\bibinfo
  {journal} {Phys. Rev.}\ }\textbf {\bibinfo {volume} {185}},\ \bibinfo {pages}
  {1576} (\bibinfo {year} {1969})}\BibitemShut {NoStop}%
\bibitem [{\citenamefont {Goryachev}\ \emph {et~al.}(1976)\citenamefont
  {Goryachev}, \citenamefont {Kuznetsov}, \citenamefont {Orlin}, \citenamefont
  {Pozhidaeva},\ and\ \citenamefont {Shevchenko}}]{Goryachev_1976}%
  \BibitemOpen
  \bibfield  {author} {\bibinfo {author} {\bibfnamefont {B.~I.}\ \bibnamefont
  {Goryachev}}, \bibinfo {author} {\bibfnamefont {Y.~V.}\ \bibnamefont
  {Kuznetsov}}, \bibinfo {author} {\bibfnamefont {V.~N.}\ \bibnamefont
  {Orlin}}, \bibinfo {author} {\bibfnamefont {N.~A.}\ \bibnamefont
  {Pozhidaeva}},\ and\ \bibinfo {author} {\bibfnamefont {V.~G.}\ \bibnamefont
  {Shevchenko}},\ }\bibfield  {title} {\bibinfo {title} {{Giant resonance in
  the strongly deformed nuclei $^{159}$Tb, $^{165}$Ho, $^{166}$Er, and
  $^{178}$Hf}},\ }\href@noop {} {\bibfield  {journal} {\bibinfo  {journal}
  {Yadernaya Fizika}\ }\textbf {\bibinfo {volume} {23}},\ \bibinfo {pages}
  {1145} (\bibinfo {year} {1976})}\BibitemShut {NoStop}%
\bibitem [{\citenamefont {Axel}\ \emph {et~al.}(1966)\citenamefont {Axel},
  \citenamefont {Miller}, \citenamefont {Schuhl}, \citenamefont {Tamas},\ and\
  \citenamefont {Tzara}}]{Axel_1966}%
  \BibitemOpen
  \bibfield  {author} {\bibinfo {author} {\bibfnamefont {P.}~\bibnamefont
  {Axel}}, \bibinfo {author} {\bibfnamefont {J.}~\bibnamefont {Miller}},
  \bibinfo {author} {\bibfnamefont {C.}~\bibnamefont {Schuhl}}, \bibinfo
  {author} {\bibfnamefont {G.}~\bibnamefont {Tamas}},\ and\ \bibinfo {author}
  {\bibfnamefont {C.}~\bibnamefont {Tzara}},\ }\bibfield  {title} {\bibinfo
  {title} {Etude de la deformation du noyau d'holmium},\ }\href
  {https://doi.org/10.1051/jphys:01966002705-6026200} {\bibfield  {journal}
  {\bibinfo  {journal} {J. Phys. France}\ }\textbf {\bibinfo {volume} {27}},\
  \bibinfo {pages} {262} (\bibinfo {year} {1966})}\BibitemShut {NoStop}%
\bibitem [{\citenamefont {Catan\u{a}}\ \emph {et~al.}(1974)\citenamefont
  {Catan\u{a}}, \citenamefont {Baciu}, \citenamefont {Ionescu-Bujor},
  \citenamefont {Niculescu},\ and\ \citenamefont {Iliescu}}]{Catana_1974}%
  \BibitemOpen
  \bibfield  {author} {\bibinfo {author} {\bibfnamefont {D.}~\bibnamefont
  {Catan\u{a}}}, \bibinfo {author} {\bibfnamefont {G.}~\bibnamefont {Baciu}},
  \bibinfo {author} {\bibfnamefont {M.}~\bibnamefont {Ionescu-Bujor}}, \bibinfo
  {author} {\bibfnamefont {V.~I.~R.}\ \bibnamefont {Niculescu}},\ and\ \bibinfo
  {author} {\bibfnamefont {C.}~\bibnamefont {Iliescu}},\ }\bibfield  {title}
  {\bibinfo {title} {{The effective cross section of the
  $^{165}$Ho($\gamma$,xn) reaction}},\ }\href
  {https://doi.org/https://doi.org/10.1016/0375-9474(74)90372-8} {\bibfield
  {journal} {\bibinfo  {journal} {Nuclear Physics A}\ }\textbf {\bibinfo
  {volume} {225}},\ \bibinfo {pages} {157} (\bibinfo {year}
  {1974})}\BibitemShut {NoStop}%
\bibitem [{\citenamefont {Gurevich}\ \emph {et~al.}(1976)\citenamefont
  {Gurevich}, \citenamefont {Lazareva}, \citenamefont {Mazur},\ and\
  \citenamefont {Solodukhov}}]{Gurevich_1976}%
  \BibitemOpen
  \bibfield  {author} {\bibinfo {author} {\bibfnamefont {G.~M.}\ \bibnamefont
  {Gurevich}}, \bibinfo {author} {\bibfnamefont {L.~E.}\ \bibnamefont
  {Lazareva}}, \bibinfo {author} {\bibfnamefont {V.~M.}\ \bibnamefont
  {Mazur}},\ and\ \bibinfo {author} {\bibfnamefont {G.~V.}\ \bibnamefont
  {Solodukhov}},\ }\bibfield  {title} {\bibinfo {title} {Width of giant
  resonance in the absorption for the cross sections of gamma-rays by nuclei in
  the region $150 < \mathrm{A} < 200$},\ }\href@noop {} {\bibfield  {journal}
  {\bibinfo  {journal} {Zhurnal Eksper. i Teoret. Fiz., Pisma v Redakt.}\
  }\textbf {\bibinfo {volume} {23}},\ \bibinfo {pages} {411} (\bibinfo {year}
  {1976})}\BibitemShut {NoStop}%
\bibitem [{\citenamefont {Gurevich}\ \emph {et~al.}(1981)\citenamefont
  {Gurevich}, \citenamefont {Lazareva}, \citenamefont {Mazur}, \citenamefont
  {Merkulov}, \citenamefont {Solodukhov},\ and\ \citenamefont
  {Tyutin}}]{Gurevich_1981}%
  \BibitemOpen
  \bibfield  {author} {\bibinfo {author} {\bibfnamefont {G.~M.}\ \bibnamefont
  {Gurevich}}, \bibinfo {author} {\bibfnamefont {L.~E.}\ \bibnamefont
  {Lazareva}}, \bibinfo {author} {\bibfnamefont {V.~M.}\ \bibnamefont {Mazur}},
  \bibinfo {author} {\bibfnamefont {S.~Y.}\ \bibnamefont {Merkulov}}, \bibinfo
  {author} {\bibfnamefont {G.~V.}\ \bibnamefont {Solodukhov}},\ and\ \bibinfo
  {author} {\bibfnamefont {V.~A.}\ \bibnamefont {Tyutin}},\ }\bibfield  {title}
  {\bibinfo {title} {Total nuclear photoabsorption cross sections in the region
  $150 < \mathrm{A} < 190$},\ }\href
  {https://doi.org/https://doi.org/10.1016/0375-9474(81)90443-7} {\bibfield
  {journal} {\bibinfo  {journal} {Nuclear Physics A}\ }\textbf {\bibinfo
  {volume} {351}},\ \bibinfo {pages} {257} (\bibinfo {year}
  {1981})}\BibitemShut {NoStop}%
\bibitem [{\citenamefont {Herman}\ \emph {et~al.}(2007)\citenamefont {Herman},
  \citenamefont {Capote}, \citenamefont {Carlson}, \citenamefont
  {Oblo\v{z}insk\'{y}}, \citenamefont {Sin}, \citenamefont {Trkov},
  \citenamefont {Wienke},\ and\ \citenamefont {Zerkin}}]{herman_2007_empire}%
  \BibitemOpen
  \bibfield  {author} {\bibinfo {author} {\bibfnamefont {M.}~\bibnamefont
  {Herman}}, \bibinfo {author} {\bibfnamefont {R.}~\bibnamefont {Capote}},
  \bibinfo {author} {\bibfnamefont {B.}~\bibnamefont {Carlson}}, \bibinfo
  {author} {\bibfnamefont {P.}~\bibnamefont {Oblo\v{z}insk\'{y}}}, \bibinfo
  {author} {\bibfnamefont {M.}~\bibnamefont {Sin}}, \bibinfo {author}
  {\bibfnamefont {A.}~\bibnamefont {Trkov}}, \bibinfo {author} {\bibfnamefont
  {H.}~\bibnamefont {Wienke}},\ and\ \bibinfo {author} {\bibfnamefont
  {V.}~\bibnamefont {Zerkin}},\ }\bibfield  {title} {\bibinfo {title} {{EMPIRE:
  Nuclear Reaction Model Code System for Data Evaluation}},\ }\href
  {https://doi.org/https://doi.org/10.1016/j.nds.2007.11.003} {\bibfield
  {journal} {\bibinfo  {journal} {Nuclear Data Sheets}\ }\textbf {\bibinfo
  {volume} {108}},\ \bibinfo {pages} {2655} (\bibinfo {year} {2007})},\
  \bibinfo {note} {$\mathrm{S}$pecial Issue on Evaluations of Neutron Cross
  Sections}\BibitemShut {NoStop}%
\bibitem [{\citenamefont {Koning}\ \emph {et~al.}(2023)\citenamefont {Koning},
  \citenamefont {Hilaire},\ and\ \citenamefont {Goriely}}]{koning_2023_talys}%
  \BibitemOpen
  \bibfield  {author} {\bibinfo {author} {\bibfnamefont {A.}~\bibnamefont
  {Koning}}, \bibinfo {author} {\bibfnamefont {S.}~\bibnamefont {Hilaire}},\
  and\ \bibinfo {author} {\bibfnamefont {S.}~\bibnamefont {Goriely}},\
  }\bibfield  {title} {\bibinfo {title} {{TALYS: modeling of nuclear
  reactions}},\ }\href {https://doi.org/10.1140/epja/s10050-023-01034-3}
  {\bibfield  {journal} {\bibinfo  {journal} {The European Physical Journal A}\
  }\textbf {\bibinfo {volume} {59}},\ \bibinfo {pages} {131} (\bibinfo {year}
  {2023})}\BibitemShut {NoStop}%
\bibitem [{\citenamefont {Capote}\ \emph {et~al.}(2009)\citenamefont {Capote},
  \citenamefont {Herman}, \citenamefont {Oblo\v{z}insk\'{y}}, \citenamefont
  {Young}, \citenamefont {Goriely}, \citenamefont {Belgya}, \citenamefont
  {Ignatyuk}, \citenamefont {Koning}, \citenamefont {Hilaire}, \citenamefont
  {Plujko}, \citenamefont {Avrigeanu}, \citenamefont {Bersillon}, \citenamefont
  {Chadwick}, \citenamefont {Fukahori}, \citenamefont {Ge}, \citenamefont
  {Han}, \citenamefont {Kailas}, \citenamefont {Kopecky}, \citenamefont
  {Maslov}, \citenamefont {Reffo}, \citenamefont {Sin}, \citenamefont
  {Soukhovitskii},\ and\ \citenamefont {Talou}}]{capote2009_ripl}%
  \BibitemOpen
  \bibfield  {author} {\bibinfo {author} {\bibfnamefont {R.}~\bibnamefont
  {Capote}}, \bibinfo {author} {\bibfnamefont {M.}~\bibnamefont {Herman}},
  \bibinfo {author} {\bibfnamefont {P.}~\bibnamefont {Oblo\v{z}insk\'{y}}},
  \bibinfo {author} {\bibfnamefont {P.}~\bibnamefont {Young}}, \bibinfo
  {author} {\bibfnamefont {S.}~\bibnamefont {Goriely}}, \bibinfo {author}
  {\bibfnamefont {T.}~\bibnamefont {Belgya}}, \bibinfo {author} {\bibfnamefont
  {A.}~\bibnamefont {Ignatyuk}}, \bibinfo {author} {\bibfnamefont
  {A.}~\bibnamefont {Koning}}, \bibinfo {author} {\bibfnamefont
  {S.}~\bibnamefont {Hilaire}}, \bibinfo {author} {\bibfnamefont
  {V.}~\bibnamefont {Plujko}}, \bibinfo {author} {\bibfnamefont
  {M.}~\bibnamefont {Avrigeanu}}, \bibinfo {author} {\bibfnamefont
  {O.}~\bibnamefont {Bersillon}}, \bibinfo {author} {\bibfnamefont
  {M.}~\bibnamefont {Chadwick}}, \bibinfo {author} {\bibfnamefont
  {T.}~\bibnamefont {Fukahori}}, \bibinfo {author} {\bibfnamefont
  {Z.}~\bibnamefont {Ge}}, \bibinfo {author} {\bibfnamefont {Y.}~\bibnamefont
  {Han}}, \bibinfo {author} {\bibfnamefont {S.}~\bibnamefont {Kailas}},
  \bibinfo {author} {\bibfnamefont {J.}~\bibnamefont {Kopecky}}, \bibinfo
  {author} {\bibfnamefont {V.}~\bibnamefont {Maslov}}, \bibinfo {author}
  {\bibfnamefont {G.}~\bibnamefont {Reffo}}, \bibinfo {author} {\bibfnamefont
  {M.}~\bibnamefont {Sin}}, \bibinfo {author} {\bibfnamefont {E.}~\bibnamefont
  {Soukhovitskii}},\ and\ \bibinfo {author} {\bibfnamefont {P.}~\bibnamefont
  {Talou}},\ }\bibfield  {title} {\bibinfo {title} {{RIPL - Reference Input
  Parameter Library for Calculation of Nuclear Reactions and Nuclear Data
  Evaluations}},\ }\href
  {https://doi.org/https://doi.org/10.1016/j.nds.2009.10.004} {\bibfield
  {journal} {\bibinfo  {journal} {Nuclear Data Sheets}\ }\textbf {\bibinfo
  {volume} {110}},\ \bibinfo {pages} {3107} (\bibinfo {year} {2009})},\
  \bibinfo {note} {special Issue on Nuclear Reaction Data}\BibitemShut
  {NoStop}%
\bibitem [{\citenamefont {Plujko}\ \emph {et~al.}(2018)\citenamefont {Plujko},
  \citenamefont {Gorbachenko}, \citenamefont {Capote},\ and\ \citenamefont
  {Dimitriou}}]{plujko_2018}%
  \BibitemOpen
  \bibfield  {author} {\bibinfo {author} {\bibfnamefont {V.~A.}\ \bibnamefont
  {Plujko}}, \bibinfo {author} {\bibfnamefont {O.~M.}\ \bibnamefont
  {Gorbachenko}}, \bibinfo {author} {\bibfnamefont {R.}~\bibnamefont
  {Capote}},\ and\ \bibinfo {author} {\bibfnamefont {P.}~\bibnamefont
  {Dimitriou}},\ }\bibfield  {title} {\bibinfo {title} {Giant dipole resonance
  parameters of ground-state photoabsorption: Experimental values with
  uncertainties},\ }\href
  {https://doi.org/https://doi.org/10.1016/j.adt.2018.03.002} {\bibfield
  {journal} {\bibinfo  {journal} {Atomic Data and Nuclear Data Tables}\
  }\textbf {\bibinfo {volume} {123-124}},\ \bibinfo {pages} {1} (\bibinfo
  {year} {2018})}\BibitemShut {NoStop}%
\bibitem [{\citenamefont {Ligensa}\ \emph {et~al.}(1966)\citenamefont
  {Ligensa}, \citenamefont {Greiner},\ and\ \citenamefont
  {Danos}}]{Ligensa_1966}%
  \BibitemOpen
  \bibfield  {author} {\bibinfo {author} {\bibfnamefont {R.}~\bibnamefont
  {Ligensa}}, \bibinfo {author} {\bibfnamefont {W.}~\bibnamefont {Greiner}},\
  and\ \bibinfo {author} {\bibfnamefont {M.}~\bibnamefont {Danos}},\ }\bibfield
   {title} {\bibinfo {title} {Nuclear giant quadrupole resonance},\ }\href
  {https://doi.org/10.1103/PhysRevLett.16.364} {\bibfield  {journal} {\bibinfo
  {journal} {Phys. Rev. Lett.}\ }\textbf {\bibinfo {volume} {16}},\ \bibinfo
  {pages} {364} (\bibinfo {year} {1966})}\BibitemShut {NoStop}%
\bibitem [{\citenamefont {Ishkhanov}\ and\ \citenamefont
  {Orlin}(2015)}]{Ishkhanov_2015}%
  \BibitemOpen
  \bibfield  {author} {\bibinfo {author} {\bibfnamefont {B.~S.}\ \bibnamefont
  {Ishkhanov}}\ and\ \bibinfo {author} {\bibfnamefont {V.~N.}\ \bibnamefont
  {Orlin}},\ }\bibfield  {title} {\bibinfo {title} {Modified version of the
  combined model of photonucleon reactions},\ }\href
  {https://doi.org/10.1134/S1063778815040067} {\bibfield  {journal} {\bibinfo
  {journal} {Physics of Atomic Nuclei}\ }\textbf {\bibinfo {volume} {78}},\
  \bibinfo {pages} {557} (\bibinfo {year} {2015})}\BibitemShut {NoStop}%
\bibitem [{kmf(2026)}]{kmfr_website}%
  \BibitemOpen
  \href@noop {} {\bibinfo {title} {$\mathrm{KMFR}$: A combined model of
  photonuclear reactions}},\ \bibinfo {howpublished}
  {http://depni.sinp.msu.ru/$\sim$hatta/kmfr} (\bibinfo {year} {2026}),\
  \bibinfo {note} {accessed: 2026-09-18}\BibitemShut {NoStop}%
\bibitem [{\citenamefont {Bohr}\ and\ \citenamefont
  {Mottelson}(1998)}]{bohr_mottelson}%
  \BibitemOpen
  \bibfield  {author} {\bibinfo {author} {\bibfnamefont {A.}~\bibnamefont
  {Bohr}}\ and\ \bibinfo {author} {\bibfnamefont {B.~R.}\ \bibnamefont
  {Mottelson}},\ }\href@noop {} {\emph {\bibinfo {title} {Nuclear Structure}}}\
  (\bibinfo  {publisher} {World Scientific},\ \bibinfo {address} {Singapore},\
  \bibinfo {year} {1998})\BibitemShut {NoStop}%
\bibitem [{\citenamefont {Angeli}\ and\ \citenamefont
  {Marinova}(2013)}]{Angeli_2013}%
  \BibitemOpen
  \bibfield  {author} {\bibinfo {author} {\bibfnamefont {I.}~\bibnamefont
  {Angeli}}\ and\ \bibinfo {author} {\bibfnamefont {K.~P.}\ \bibnamefont
  {Marinova}},\ }\bibfield  {title} {\bibinfo {title} {{Table of experimental
  nuclear ground state charge radii: An update}},\ }\href
  {https://doi.org/https://doi.org/10.1016/j.adt.2011.12.006} {\bibfield
  {journal} {\bibinfo  {journal} {Atomic Data and Nuclear Data Tables}\
  }\textbf {\bibinfo {volume} {99}},\ \bibinfo {pages} {69} (\bibinfo {year}
  {2013})}\BibitemShut {NoStop}%
\bibitem [{\citenamefont {Olaniyi}\ \emph {et~al.}(1982)\citenamefont
  {Olaniyi}, \citenamefont {Shor}, \citenamefont {Cheng}, \citenamefont
  {Dugan},\ and\ \citenamefont {Wu}}]{Olaniyi_1982}%
  \BibitemOpen
  \bibfield  {author} {\bibinfo {author} {\bibfnamefont {B.}~\bibnamefont
  {Olaniyi}}, \bibinfo {author} {\bibfnamefont {A.}~\bibnamefont {Shor}},
  \bibinfo {author} {\bibfnamefont {S.~C.}\ \bibnamefont {Cheng}}, \bibinfo
  {author} {\bibfnamefont {G.}~\bibnamefont {Dugan}},\ and\ \bibinfo {author}
  {\bibfnamefont {C.~S.}\ \bibnamefont {Wu}},\ }\bibfield  {title} {\bibinfo
  {title} {Electric quadrupole moments and strong interaction effects in pionic
  atoms of $^{165}\text{Ho}$, $^{175}\text{Lu}$, $^{176}\text{Lu}$,
  $^{179}\text{Hf}$ and $^{181}\text{Ta}$},\ }\href
  {https://doi.org/10.1016/0375-9474(83)90623-1} {\bibfield  {journal}
  {\bibinfo  {journal} {Nuclear Physics A}\ }\textbf {\bibinfo {volume}
  {403}},\ \bibinfo {pages} {572} (\bibinfo {year} {1982})}\BibitemShut
  {NoStop}%
\bibitem [{\citenamefont {Stone}(2021)}]{stone_2021}%
  \BibitemOpen
  \bibfield  {author} {\bibinfo {author} {\bibfnamefont {N.~J.}\ \bibnamefont
  {Stone}},\ }\href {https://doi.org/10.61092/iaea.a6te-dg7q} {\emph {\bibinfo
  {title} {Table of Nuclear Electric Quadrupole Moments}}},\ \bibinfo {type}
  {Tech. Rep.}\ \bibinfo {number} {INDC(NDS)-0833}\ (\bibinfo  {institution}
  {IAEA Nuclear Data Section},\ \bibinfo {address} {Vienna, Austria},\ \bibinfo
  {year} {2021})\BibitemShut {NoStop}%
\bibitem [{\citenamefont {Capote}\ \emph {et~al.}(1991)\citenamefont {Capote},
  \citenamefont {Osorio}, \citenamefont {L{\'o}pez}, \citenamefont {Herrera},\
  and\ \citenamefont {Piris}}]{capote_1991_pcross}%
  \BibitemOpen
  \bibfield  {author} {\bibinfo {author} {\bibfnamefont {R.}~\bibnamefont
  {Capote}}, \bibinfo {author} {\bibfnamefont {V.}~\bibnamefont {Osorio}},
  \bibinfo {author} {\bibfnamefont {R.}~\bibnamefont {L{\'o}pez}}, \bibinfo
  {author} {\bibfnamefont {E.}~\bibnamefont {Herrera}},\ and\ \bibinfo {author}
  {\bibfnamefont {M.}~\bibnamefont {Piris}},\ }\href@noop {} {\emph {\bibinfo
  {title} {Analysis of experimental data on neutron-induced reactions and
  development of code {PCROSS} for the calculation of differential
  pre-equilibrium emission spectra with modelling of the level density
  function}}},\ \bibinfo {type} {Tech. Rep.}\ \bibinfo {number}
  {IAEA(CUB)-004}\ (\bibinfo  {institution} {International Atomic Energy
  Agency},\ \bibinfo {address} {Vienna, Austria},\ \bibinfo {year} {1991})\
  \bibinfo {note} {{Final report on research contract 5472/RB}}\BibitemShut
  {NoStop}%
\bibitem [{\citenamefont {Orlin}\ and\ \citenamefont
  {Stopani}(2023)}]{Orlin_2023}%
  \BibitemOpen
  \bibfield  {author} {\bibinfo {author} {\bibfnamefont {V.~N.}\ \bibnamefont
  {Orlin}}\ and\ \bibinfo {author} {\bibfnamefont {K.~A.}\ \bibnamefont
  {Stopani}},\ }\bibfield  {title} {\bibinfo {title} {{Direct photoeffect in
  heavy deformed nuclei at $E_{\gamma} \le 40$~\text{MeV}}},\ }\href
  {https://doi.org/10.1103/PhysRevC.108.054606} {\bibfield  {journal} {\bibinfo
   {journal} {Physical Review C}\ }\textbf {\bibinfo {volume} {108}},\ \bibinfo
  {pages} {054606} (\bibinfo {year} {2023})}\BibitemShut {NoStop}%
\bibitem [{\citenamefont {Goriely}\ \emph {et~al.}(2008)\citenamefont
  {Goriely}, \citenamefont {Hilaire},\ and\ \citenamefont
  {Koning}}]{goriely_2008}%
  \BibitemOpen
  \bibfield  {author} {\bibinfo {author} {\bibfnamefont {S.}~\bibnamefont
  {Goriely}}, \bibinfo {author} {\bibfnamefont {S.}~\bibnamefont {Hilaire}},\
  and\ \bibinfo {author} {\bibfnamefont {A.~J.}\ \bibnamefont {Koning}},\
  }\bibfield  {title} {\bibinfo {title} {Improved microscopic nuclear level
  densities within the {HFB} plus combinatorial method},\ }\href
  {https://doi.org/10.1103/PhysRevC.78.064307} {\bibfield  {journal} {\bibinfo
  {journal} {Physical Review C}\ }\textbf {\bibinfo {volume} {78}},\ \bibinfo
  {pages} {064307} (\bibinfo {year} {2008})}\BibitemShut {NoStop}%
\bibitem [{\citenamefont {Hilaire}\ \emph {et~al.}(2012)\citenamefont
  {Hilaire}, \citenamefont {Girod}, \citenamefont {Goriely},\ and\
  \citenamefont {Koning}}]{hilaire_2012}%
  \BibitemOpen
  \bibfield  {author} {\bibinfo {author} {\bibfnamefont {S.}~\bibnamefont
  {Hilaire}}, \bibinfo {author} {\bibfnamefont {M.}~\bibnamefont {Girod}},
  \bibinfo {author} {\bibfnamefont {S.}~\bibnamefont {Goriely}},\ and\ \bibinfo
  {author} {\bibfnamefont {A.~J.}\ \bibnamefont {Koning}},\ }\bibfield  {title}
  {\bibinfo {title} {Temperature-dependent combinatorial level densities with
  the {D1M} {Gogny} force},\ }\href
  {https://doi.org/10.1103/PhysRevC.86.064317} {\bibfield  {journal} {\bibinfo
  {journal} {Physical Review C}\ }\textbf {\bibinfo {volume} {86}},\ \bibinfo
  {pages} {064317} (\bibinfo {year} {2012})}\BibitemShut {NoStop}%
\bibitem [{\citenamefont {Goriely}\ \emph {et~al.}(2026)\citenamefont
  {Goriely}, \citenamefont {Ryssens}, \citenamefont {Hilaire},\ and\
  \citenamefont {Koning}}]{goriely_2026}%
  \BibitemOpen
  \bibfield  {author} {\bibinfo {author} {\bibfnamefont {S.}~\bibnamefont
  {Goriely}}, \bibinfo {author} {\bibfnamefont {W.}~\bibnamefont {Ryssens}},
  \bibinfo {author} {\bibfnamefont {S.}~\bibnamefont {Hilaire}},\ and\ \bibinfo
  {author} {\bibfnamefont {A.~J.}\ \bibnamefont {Koning}},\ }\bibfield  {title}
  {\bibinfo {title} {Improved microscopic nuclear level densities within the
  triaxial {Hartree}-{Fock}-{Bogoliubov} plus combinatorial method},\ }\href
  {https://doi.org/10.1103/PhysRevC.113.014320} {\bibfield  {journal} {\bibinfo
   {journal} {Physical Review C}\ }\textbf {\bibinfo {volume} {113}},\ \bibinfo
  {pages} {014320} (\bibinfo {year} {2026})}\BibitemShut {NoStop}%
\bibitem [{sup()}]{supplemental_material}%
  \BibitemOpen
  \href@noop {} {}\bibinfo {note} {See Supplemental Material at [URL will be
  inserted by publisher] for the $^{165}\text{Ho}$ and $^{169}\text{Tm}$
  experimental data points: the $(\gamma,\,S\mathrm{n})$ cross sections, and
  the $(\gamma,\,1nX)$, $(\gamma,\,2nX)$, $(\gamma,\,3nX)$ and $(\gamma,\,4nX)$
  cross sections and average photoneutron energies.}\BibitemShut {Stop}%
\end{thebibliography}%

\end{document}